\documentclass[11pt]{article}
\pdfoutput=1
\usepackage{setspace, graphicx, amssymb, amsmath, latexsym, amsfonts, amscd, amsthm, multirow, mathdots,caption,array,diagbox,mathtools}
\usepackage{tabularx,cite,mathrsfs,tikz,subfig}
\usepackage{authblk}
\usepackage{color}

\usepackage[margin=2cm]{caption}
\usepackage{amsmath}
\usepackage{fullpage}
\usepackage{stmaryrd}
\usepackage{rotating}

\usepackage{hyperref}

\def\bR {\mathbb{R}}
\def\bZ {\mathbb{Z}}

\newcommand{\cM}{{\cal M }}

\newcommand{\cL}{{\cal L }}                
\newcommand{\cO}{{\cal O }}            
\newcommand{\cH}{{\cal H }}

\newcommand{\be}{\begin{equation}}
\newcommand{\ee}{\end{equation}}
\newcommand{\bea}{\begin{eqnarray}}
\newcommand{\eea}{\end{eqnarray}}
\newcommand{\nn}{\nonumber}

\newcommand{\mc}{\mathcal}
\def\({\left(}
\def\){\right)}

\newcommand{\e}{\epsilon}

\newcommand{\Tr}{{\rm Tr \,}}

\numberwithin{equation}{section}
\def\ie{\begin{equation}\begin{aligned}}
\def\fe{\end{aligned}\end{equation}}

\newcommand{\Z}{\mathbb{Z}}

\newcommand{\tf}{\textstyle\frac}

\begin{document}

\begin{titlepage}
	\begin{center}
		
		\hfill \\
		\hfill \\
		\vskip 0.75in
		
		{\Huge
			\bf Narain to Nar\textcolor{gray}{n}i\textcolor{gray}{a}
		}\\

		\vskip 0.7in
		
		{\large Nathan Benjamin,$^{a}$ Christoph A.~Keller,${}^{b}$ Hirosi  Ooguri,${}^{c,d}$ and Ida G.~Zadeh${}^{e}$
		}\\
		\vskip 0.4in
		
	${}^a$
{\it Princeton Center for Theoretical Science, Princeton University, 
Princeton, NJ 08544, USA}\vskip 1mm
		${}^{b}$
		{\it Department of Mathematics, University of Arizona, Tucson, AZ 85721-0089, USA} \vskip 1mm
		${}^{c}$
		{\it Walter Burke Institute for Theoretical Physics, \\ California Institute of Technology, Pasadena, CA 91125, USA} \vskip 1mm	
		${}^{d}$
		{\it Kavli Institute for the Physics and Mathematics of the Universe (WPI),\\ University of Tokyo, Kashiwa, 277-8583, Japan		
		}	\vskip 1mm	
		${}^{e}$
		{\it International Centre for Theoretical Physics, Strada Costiera 11, 34151 Trieste, Italy} 

\vskip 0.2in

		\texttt{nathanb\_at\_princeton.edu, cakeller\_at\_math.arizona.edu, \\ ooguri\_at\_caltech.edu, zadeh\_at\_ictp.it}

	\end{center}
	
	\vskip 0.35in
	
	\begin{center} {\bf ABSTRACT } \end{center}
We generalize the holographic correspondence between 
topological gravity coupled to an abelian Chern-Simons theory in three dimensions and 
 an ensemble average of Narain's family of massless free bosons in two dimensions, discovered
by Afkhami-Jeddi {\it et al.} and by Maloney and Witten.
We find that the correspondence also works for toroidal orbifolds but not for K3 or Calabi-Yau sigma-models and
not always for the minimal models. We conjecture that the correspondence requires that the
central charge is equal to the critical central charge defined by the asymptotic
density of states of the chiral algebra. For toroidal orbifolds, we extend the holographic 
correspondence to correlation functions of twist operators by using
topological properties of rational tangles in the 
three-dimensional ball, which represent configurations of vortices associated to a discrete gauge symmetry. 

	\vfill
	
	\noindent {\tt v1:} March 29, 2021 \\ {\tt v2:} September 7, 2021

\end{titlepage}
\tableofcontents
\section{Introduction}
Recently, a new type of holographic pair has been found, in which quantum gravity theories
are dual to random ensembles of quantum mechanical systems in one and
two dimensions. In \cite{Saad:2019lba}, an average of a certain set of
quantum mechanical systems is found to be dual to Jackiw-Teitelboim gravity 
in two dimensions. In \cite{Afkhami-Jeddi:2020ezh, Maloney:2020nni}, the average of
partition functions of $c$ massless free bosons in two dimensions 
over Narain moduli space 
 is shown to be equal to that
of a $U(1)^c \times U(1)^c$ Chern-Simons gauge theory in three dimensions
coupled to topological gravity, which generates a sum over three-dimensional 
topologies allowing hyperbolic metrics. 
These discoveries have revived questions
on the role of wormholes in quantum gravity \cite{Coleman:1988cy, Coleman:1988tj} and its implications in holography \cite{Maldacena:2004rf}. 
 Unlike traditional examples of
the AdS/CFT correspondence derived from string theory, there is no known
fundamental explanation for the emergence of the ensemble averages in these correspondences. 
It is desirable to have more examples so that we can identify
general patterns in holography with ensemble averages
 and learn lessons on wormholes and sums over topologies in quantum gravity
in general. 

In this paper, we examine three more groups of candidates for holographic 
conformal field theories (CFTs) in two dimensions. They are $\mathbb Z_N$ orbifolds of massless
free boson theories, superconformal sigma-models with Calabi-Yau target spaces, and the minimal models. 
 We find the holographic correspondence with an ensemble average works for 
the orbifold models, but not for the Calabi-Yau sigma-models and not always for the minimal models. 
This leads us to conjecture 
that the correspondence requires that the
central charge $c$ is equal to the critical central charge  $c_{\rm crit}$
defined by the asymptotic density of states of the chiral algebra. 
For the case of the Calabi-Yau sigma-models,  $c$ is $3$ times
the complex dimension of the target Calabi-Yau manifold while $c_{\rm crit}=3$, and thus $c > c_{\rm crit}$ except for the $2$-torus, where the correspondence works.
 On the other hand, $c = c_{\rm crit}$ for the
orbifold models and the minimal models.

In \cite{Afkhami-Jeddi:2020ezh, Maloney:2020nni}, the correspondence between a Chern-Simons theory and the averaged massless free boson theory was tested for their partition 
functions, but not for their correlation functions. 
In fact, there seems to be no non-trivial correspondence for correlation
functions in this case. Correlation functions of 
the $U(1)^c \times U(1)^c$ currents in the free boson theory
do not depend on the Narain moduli, and their ensemble averages are trivial.  
While correlation functions of operators with non-zero momenta
and winding numbers have non-trivial dependence on the Narain moduli,
there are no observables in the bulk corresponding to their averages since 
there are no localized states carrying non-zero charges with respect to the $U(1)^c \times U(1)^c$ gauge
symmetry.

If we orbifoldize the free boson theory, a non-trivial duality can be found for 
correlation functions. The bulk theory dual to the $\mathbb Z_N$ orbifold
is a pure Chern-Simons gauge theory with its $\mathbb Z_N$ global symmetry gauged, 
 coupled to topological gravity. We will show that the averages of twist operator correlation functions in the CFT are equal to sums of correlation functions of vortices associated to the $\mathbb Z_N$ 
gauge symmetry in the bulk.
These sums are over configurations of the vortices ending on the twist operators on the boundary,
and are restricted to those called
{\it rational tangles} in three-dimensional knot theory \cite{Conway}.
It turns out that the $\mathbb Z_N$ branched cover over a tangle configuration is
a genus-$(N-1)$ handlebody if and only if the tangle is rational. Therefore, the sum is restricted to vortex configurations
whose $N$-fold branched covers allow hyperbolic metrics.\footnote{Note, however, 
the subtlety for $N > 2$ due to the existence of non-handlebodies with 
hyperbolic metrics \cite{Yin:2007at}, 
which is also an issue with the higher genus partition functions in the $T^c$ case \cite{Afkhami-Jeddi:2020ezh, Maloney:2020nni}.} 

The rest of the paper is organized as follows. In Section \ref{sec:ReviewNarain}, we review the Narain ensemble average of \cite{Afkhami-Jeddi:2020ezh, Maloney:2020nni}. In Section \ref{sec:tcz2}, we generalize the ensemble average to $\mathbb{Z}_2$ orbifolds. 
In Section \ref{sec:tczn}, we generalize the ensemble average to a class of $\mathbb{Z}_N$ orbifolds. In Section \ref{s:corrfcts}, we calculate the average correlation functions in the
$\mathbb{Z}_2$ orbifold case  and give a bulk interpretation. In Section \ref{sec:ZNCor}, we do the same for the $\mathbb{Z}_N$ orbifolds. 
In Section \ref{sec:Nonfac}, we briefly discuss some puzzles about non-factorizable amplitudes.
In Section \ref{sec:K3Gen}, we discuss ensemble averages of
 K3 and Calabi-Yau sigma models. Some detailed calculations are left to the appendices: In Appendix \ref{app:PoinVac}, we compute the Poincar\'e sum of the $U(1)^c/\mathbb{Z}_2$ vacuum character. In Appendix \ref{app:Kahler}, we derive a differential equation the theta functions of $\mathbb{Z}_N$ orbifold theories obey. In Appendices \ref{sec:c6kernels} and \ref{app:mwkcy}, we derive modular kernels for the $\mathcal{N}=2, 4$ superconformal algebras. In Appendix \ref{app:minmod}, we compute Poincar\'e sums of minimal model vacuum characters.

\bigskip
\noindent
{\bf Note on the title of this paper}: 

\noindent
The title is in homage to the  video introduction to the study of branched coverings over 
knots entitled  ``Knots to Narnia" by W. Thurston \cite{Thurston}, 
which will play an important role in this paper.
We thank Tom Melia  for pointing out that Narnia is an anagram of Narain.

\bigskip

\section{Review of Narain averaging duality}
\label{sec:ReviewNarain}

In this section we review the recent work \cite{Afkhami-Jeddi:2020ezh, Maloney:2020nni} which established a holographic duality between an average of $c$ free bosons and an exotic Chern-Simons-like theory of gravity in three dimensions. Consider a CFT of $c$ free bosons with $c>2$. The parameter space of this theory is the $c^2$-dimensional Narain moduli space
\be
O(c,c;\mathbb Z) \backslash O(c,c) / O(c)\times O(c)\ .
\label{eq:occ}
\ee
An averaging procedure using the Zamolodchikov measure allows one to define an ``averaged partition function." If $c>1$, the volume of this moduli space under the Zamolodchikov measure is finite, and if $c>2$ the average is also finite. To calculate the average, we review the logic in \cite[section 2.2]{Maloney:2020nni}. Let us rewrite the partition function of $c$ free bosons as
\be
Z_{m}(\tau,\bar\tau) = \frac{\Theta_m(\tau,\bar\tau)}{|\eta(\tau)|^{2c}}\ ,
\label{eq:thetadef}
\ee
where $\Theta_m(\tau,\bar\tau)$ is the Siegel-Narain theta function which contains the lattice sum --- see eqs. (\ref{thetacpx})-(\ref{latsumQ_ii}) for definitions --- and $m$ is an abstract coordinate on the target space of $c$ free bosons which parametrizes the symmetric and anti-symmetric matrices $G_{mn}, B_{mn}$.

The only moduli-dependence in eq. (\ref{eq:thetadef}) is in the theta function $\Theta_m(\tau,\bar\tau)$. A direct calculation shows that this function obeys the differential equation
\be
\Big(\Delta_{\mathcal{H}} - c\tau_2\frac{\partial}{\partial\tau_2} - \Delta_{\mathcal{M}}\Big)\Theta_m(\tau,\bar\tau) = 0
\ee
where $\Delta_{\mathcal{H}}$ is the Laplacian on the upper half plane (worldsheet moduli space), and $\Delta_{\mathcal{M}}$ is the Laplacian on (\ref{eq:occ}) (target space moduli space). Integrating over the target space moduli space and using integration by parts, \cite{Maloney:2020nni} finally gets
\be
\(\Delta_{\mathcal{H}} + \frac{c}2\(\frac{c}2 - 1\)\) \(\tau_2^{c/2}\langle\Theta_m(\tau,\bar\tau)\rangle\)=0\ ,
\label{eq:LaplacianWeWant}
\ee
where the averaging $\langle \ldots \rangle$ is over the target space coordinate $m$. Using the growth properties of $\big(\tau_2^{c/2}\langle\Theta_m(\tau,\bar\tau)\rangle\big)$, it can then be shown that eq. (\ref{eq:LaplacianWeWant}) has as unique solution given by the \emph{real analytic Eisenstein series} $E(s,\tau,\bar\tau)$:
\be\label{eisen}
\tau_2^{c/2}\langle\Theta_m(\tau,\bar\tau)\rangle = E\(\frac{c}2, \tau,\bar\tau\) \ ,
\qquad E(s,\tau,\bar\tau) =\!\!\!\!\!\! \sum_{\gamma\in\Gamma_\infty \backslash SL(2,\mathbb Z)} \!\!\!{\gamma(\tau_2)}^s\ ,
\ee
where $\Gamma_\infty$ is the group generated by the modular $T$ transformation.

This result is a rewrite of an argument originally by Siegel, and is known as the Siegel-Weil formula. The final expression for the average partition function then reads
\be
\langle Z_m(\tau,\bar\tau) \rangle = \frac{E\(\frac{c}{2},\tau,\bar\tau\)}{\tau_2^{c/2}|\eta(\tau)|^{2c}}\ .
\label{eq:AverageZME}
\ee
By using modular properties of $\eta(\tau)$, eq. (\ref{eq:AverageZME}) can be rewritten as
\be
\langle Z_m(\tau,\bar\tau) \rangle = \sum_{\gamma\in\Gamma_\infty \backslash SL(2,\mathbb Z)}\frac{1}{|\eta(\gamma\tau)|^{2c}}\ . 
\label{eq:PoincareFirstTime}
\ee
The sum (\ref{eq:PoincareFirstTime}) is known as a Poincar\'e series and is related to a sum over classical geometries in AdS$_3$. There have been various attempts in interpreting Poincar\'e series as a sum over classical saddles in computing a path integral in AdS$_3$ \cite{Dijkgraaf:2000fq, Manschot:2007ha, Maloney:2007ud, Keller:2014xba}. This then leads to a natural question: can the sum (\ref{eq:PoincareFirstTime}) be interpreted as a sum over geometries weighted by a classical action? In \cite{Afkhami-Jeddi:2020ezh, Maloney:2020nni} it was argued that it can. However, the action is not that of an Einstein-Hilbert term, but rather a $U(1)^c \times U(1)^c$ Chern-Simons theory. The theory is then coupled to a topological gravity term which induces a sum over geometries. The contribution of the Chern-Simons action to the path integral in thermal AdS$_3$ is $1/|\eta(\tau)|^{2c}$, and the sum over geometries gives us the Poincar\'e sum in (\ref{eq:PoincareFirstTime}). 

To summarize, the result of \cite{Afkhami-Jeddi:2020ezh, Maloney:2020nni} is that the following three quantities are the same:
\begin{enumerate}
\item The average partition function of $c$ free bosons.
\be
\langle Z\rangle = \frac{\int_{\mathcal{M}} d\mu Z(\mu)}{\int_{\mathcal{M}} d\mu}= \frac{E\(\frac{c}{2},\tau,\bar\tau\)}{\tau_2^{c/2}|\eta(\tau)|^{2c}}\ ,
~~~~~~ \mathcal{M} = O(c,c;\mathbb Z) \backslash O(c,c) / O(c)\times O(c)\ .
\ee
\item The Poincar\'e sum of a $U(1)^c$ vacuum character.
\be
Z = \!\!\!\!\! \sum_{\gamma \in \Gamma_\infty \backslash SL(2,\mathbb Z)}\!\!\!\!\! 
 |\chi^{\text{vac}}(\gamma\tau)|^2= \frac{E\(\frac{c}{2},\tau,\bar\tau\)}{\tau_2^{c/2}|\eta(\tau)|^{2c}}\ ,
~~~~~~~\chi^{\text{vac}}(\tau) = \frac{1}{\eta(\tau)^c}\ .
\ee
\item An exotic 3d gravity computation of a sum over geometries of a $U(1)^c \times U(1)^c$ abelian Chern-Simons theory:
\be
Z =\!\!\!\!\! \sum_{\text{3-manifold geometries}} \!\!\!\!\!\!\!\!\!\!e^{-S_{\text{CS}}}= \frac{E\(\frac{c}{2},\tau,\bar\tau\)}{\tau_2^{c/2}|\eta(\tau)|^{2c}}\ .
\ee
\end{enumerate}

\section{Averaging over $T^c/\mathbb Z_2$}
\label{sec:tcz2}

In this section we will first generalize the computation of the average partition function of $c$ free bosons to that of a $\mathbb{Z}_2$ orbifold of $c$ free bosons. We will then provide a bulk interpretation of this average. 

\subsection{Boundary CFT}
Let $Z_{T^c}(\tau,\bar\tau)$ be the partition function of the sigma-model with target space $T^c$:
\be\label{Z3Gamma11}
Z_{T^c}(\tau,\bar\tau)=\frac1{|\eta(\tau)|^{2c}}\sum_{p_L,p_R\in\Gamma_{c,c}}q^{\frac12 p_L^2}\bar q^{\frac12 p_R^2}\ ,
\ee
where $c$ is the complex dimension of the torus and $\Gamma_{c,c}$ is the Narain lattice. The characters of the $T^c/\mathbb{Z}_2$ orbifold are given by the following. There are four discrete representations as well as a continuous family. The discrete representations are \cite{DiFrancesco:1997nk}:
\begin{align}
\chi^{\text{vac}}(\tau) &= \frac12\left[\frac1{\eta(\tau)^c} + \frac{\eta(\tau)^c}{\eta(2\tau)^c}\right] = \frac12\left[\frac1{\eta(\tau)^c} + \frac{2^{\frac c2} \eta(\tau)^{\frac c2}}{\theta_2(\tau)^{\frac c2}} \right]\ , \nn\\
\chi^1(\tau) &= \frac12\left[\frac1{\eta(\tau)^c} - \frac{\eta(\tau)^c}{\eta(2\tau)^c}\right] = \frac12\left[\frac1{\eta(\tau)^c} - \frac{2^{\frac c2} \eta(\tau)^{\frac c2}}{\theta_2(\tau)^{\frac c2}} \right]\ , \nn\\
\chi^2(\tau) &= \frac12 \left[\frac{\eta(\tau)^{\frac c2}}{\theta_4(\tau)^{\frac c2}} + \frac{\eta(\tau)^{\frac c2}}{\theta_3(\tau)^{\frac c2}} \right]\ , \nn\\
\chi^3(\tau) &= \frac12 \left[\frac{\eta(\tau)^{\frac c2}}{\theta_4(\tau)^{\frac c2}} - \frac{\eta(\tau)^{\frac c2}}{\theta_3(\tau)^{\frac c2}} \right]\ .
\label{eq:CharOrb}
\end{align}
These characters have weight $0$, $1$, $\frac c{16}$, and $\frac{c}{16}+\frac12$, respectively. 
Finally, there is a family of characters with conformal weight $h$:
\be
\chi^h(\tau) = \frac{q^{h}}{\eta(\tau)^c}\ .
\label{eq:charCont}
\ee
$\chi^{\text{vac}}$ and $\chi^h$ are characters of states in the untwisted sectors that are invariant under the (left-moving) $\Z_2$ orbifold action, and $\chi^1$ is the character of the $\Z_2$-odd untwisted sector states; $\chi^2$ and $\chi^3$ are the $\Z_2$ even and odd states in the twisted sector.
The partition function is then obtained by combining states of the same left- and right-moving $\Z_2$ parity, so that they are invariant under the diagonal $\Z_2$ action. It is given by \cite{DiFrancesco:1997nk}
\be
Z_{T^c/\mathbb Z_2}(\tau,\bar\tau) = |\chi^{\text{vac}}(\tau)|^2 + |\chi^1(\tau)|^2 + 
2^c |\chi^2(\tau)|^2 + 2^c |\chi^3(\tau)|^2 + \sum_{h, \bar h} \chi^h(\tau) \chi^{\bar h}(-\bar\tau)
\label{eq:PFDecompChars}
\ee
where the last term in (\ref{eq:PFDecompChars}) is the momentum and winding sum, generically with degeneracy $1$ (unlike the $T^c$ theory where the degeneracy is generically 2 due to the $(n, m) \leftrightarrow (-n,-m)$ symmety). 
Using eq. (\ref{Z3Gamma11}) and the explicit expressions in eq. (\ref{eq:CharOrb}), we can write the partition function (\ref{eq:PFDecompChars}) as:
\be
Z_{T^c/\mathbb Z_2}(\tau, \bar \tau) = \frac12\(Z_{T^c}(\tau, \bar \tau)+
2^c\left[\Big|\frac{\eta(\tau)}{\theta_2(\tau)}\Big|^c + \Big|\frac{\eta(\tau)}{\theta_3(\tau)}\Big|^c + \Big|\frac{\eta(\tau)}{\theta_4(\tau)}\Big|^c\right]\)\ .
\label{eq:PFTdZ2}
\ee

Given a function $f(\tau,\bar\tau)$ obeying $f(\tau, \bar\tau) = f(\tau+1, \bar\tau+1)$, we define its regularized Poincar\'e series as
\be\label{Podef}
\sum_{\gamma\in \Gamma_\infty \backslash SL(2,\mathbb Z)} f(\gamma\tau, \gamma\bar\tau) = \bigg(\sum_{\gamma \in \Gamma_\infty \backslash SL(2,\mathbb Z)} (\text{Im}(\tau))^s f(\gamma\tau,\gamma\bar\tau) \bigg)_{s=0}\,
\ee
where $\Gamma_\infty$ is the group generated by the $T$ transformation (this group is isomorphic to $\mathbb Z$). 
This means that we pick a complex parameter $s$ whose real part is large enough such that the sum in (\ref{Podef}) converges, and then analytically continue the resulting function of $s$ to $s=0$.
A main result of \cite{Afkhami-Jeddi:2020ezh, Maloney:2020nni} is that for a torus CFT:
\be\label{AJM}
\sum_{\gamma\in\Gamma_\infty \backslash SL(2,\mathbb Z)} \(\frac{1}{|\eta(\gamma\tau)|^{2c}}\) = \langle Z_{T^c} \rangle_{\cM}\ .
\ee
Our goal is to generalize this to toroidal orbifold theories.

First note that the torus CFT has the $\Z_2$ symmetry acting as $\vec X \mapsto -\vec X$ everywhere on the moduli space $\cM$. Moreover, the shared chiral algebra for the orbifold is given by the currents shared at all points in the moduli space of (\ref{eq:PFTdZ2}). This is given by the first two terms (the last two terms have a ground state energy coming from the twisted sector). Using the identity
\be
\theta_2(\tau) = \frac{2\eta(2\tau)^2}{\eta(\tau)}
\ee
we can show that the shared chiral algebra for (\ref{eq:PFTdZ2}) is of the form
\be
\chi^{\text{vac}}(\tau) = \frac12\left[\frac1{\eta(\tau)^c} + \frac{\eta(\tau)^c}{\eta(2\tau)^c}\right]\ .
\label{eq:VacCharS1Z2}
\ee

It is straightforward to compute $\langle Z_{T^c/\mathbb Z_2} \rangle_{\cM}$ using eq. (\ref{eq:PFTdZ2}).
The first term on the \textsc{rhs} of eq. (\ref{eq:PFTdZ2}) is the same as the partition function of the unorbifolded theory. Moreover, the Zamolodchikov measure on the $\mathbb{Z}_2$ branch of the moduli space is the same as the unorbifolded branch, because the exactly marginal operator is unaffected by the orbifold projection. We therefore have from \cite{Afkhami-Jeddi:2020ezh, Maloney:2020nni}:
\be
\langle Z_{T^c}(\tau, \bar\tau) \rangle_{\cM} = \frac{E\(\frac c2, \tau, \bar\tau\)}{\tau_2^\frac{c}2 |\eta(\tau)|^{2c}}\ .
\label{eq:langles}
\ee
Since the remaining terms in (\ref{eq:PFTdZ2}) are moduli-independent, we find
\be
\langle Z_{T^c/\mathbb Z_2}(\tau,\bar\tau) \rangle_{\cM} = \frac12\(\frac{E\(\frac c2, \tau, \bar\tau\)}{\tau_2^\frac{c}2 |\eta(\tau)|^{2c}}+ 2^c\left[\Big|\frac{\eta(\tau)}{\theta_2(\tau)}\Big|^c + \Big|\frac{\eta(\tau)}{\theta_3(\tau)}\Big|^c + \Big|\frac{\eta(\tau)}{\theta_4(\tau)}\Big|^c\right]\).
\label{eq:avgorbz2}
\ee

Next let us generalize the \textsc{lhs} of eq. (\ref{AJM}). A naive guess might be to again take the Poincar\'e series of the vacuum contribution, namely
\be
\sum_{\gamma\in\Gamma_\infty\backslash SL(2,\mathbb Z)} |\chi^{\text{vac}}(\gamma\tau)|^2\ ,
\ee
where $\chi^{\text{vac}}(\tau)$ given by eq. (\ref{eq:VacCharS1Z2}). This turns out to be an incorrect choice: a careful computation shows that it does not agree with the average (\ref{eq:avgorbz2}) --- see appendix~\ref{app:PoinVac} for the details.

The correct prescription turns out to be the following: define $Z^{\text{st}}$ to be the contributions of all moduli-independent pieces. In the original torus case we had $Z^{\text{st}}=\frac{1}{|\eta(\tau)|^{2c}}$, since all the primaries were moduli-dependent. In our case, however, we also have the moduli-independent twisted sectors, so that
\be\label{Zin}
Z^{\text{st}}(\tau,\bar\tau) = |\chi^{\text{vac}}(\tau)|^2 + |\chi^1(\tau)|^2 + 2^c |\chi^2(\tau)|^2 + 2^c |\chi^3(\tau)^2|\ .
\ee
Here $Z^{\text{st}}$ is the moduli-independent pieces of the CFT, but we will see later in subsection \ref{ss:CSZ2} that it is also the partition function of some bulk theory on the solid torus. (In $Z^{\text{st}}$, st stands for ``solid torus".) The modular sum of $Z^{\text{st}}$ is given by
\begin{align}
&\sum_{\gamma\in\Gamma_\infty \backslash SL(2,\mathbb Z)}(Z^{\text{st}}(\gamma\tau,\gamma\bar\tau)) = \sum_{\gamma \in \Gamma_\infty \backslash SL(2,\mathbb Z)} |\chi^{\text{vac}}(\gamma\tau)|^2 + |\chi^{1}(\gamma\tau)|^2 + 2^c|\chi^{2}(\gamma\tau)|^2 + 2^c|\chi^{3}(\gamma\tau)|^2  \nn\\
&\qquad\qquad=  \frac1{2} \sum_{\gamma \in \Gamma_\infty \backslash SL(2,\mathbb Z)} \frac1{|\eta(\gamma\tau)|^{2c}} + \frac12 \sum_{\gamma \in \Gamma_\infty \backslash SL(2,\mathbb Z)} 2^c \left[ \Big | \frac{\eta(\gamma\tau)}{\theta_2(\gamma\tau)} \Big |^c +\Big | \frac{\eta(\gamma\tau)}{\theta_3(\gamma\tau)} \Big |^c +\Big | \frac{\eta(\gamma\tau)}{\theta_4(\gamma\tau)} \Big |^c \right].
\label{eq:NewMWKFixed}
\end{align}
Note that the second term in the second line of eq. (\ref{eq:NewMWKFixed}) is modular invariant so we can pull it out of the sum and be left with the divergent sum over $SL(2,\mathbb Z)$ orbits. We take this to be the Eisenstein series analytically continued to $s=0$ which gives $1$, so that formally we have
\be
\sum_{\gamma\in\Gamma_\infty\backslash SL(2,\mathbb Z)} 1 = 1\ .
\ee
The first term in eq. (\ref{eq:NewMWKFixed}) is the calculation done in \cite{Afkhami-Jeddi:2020ezh, Maloney:2020nni} and we finally get
\be\label{Z2Poincare}
\sum_{\gamma\in\Gamma_\infty \backslash SL(2,\mathbb Z)}(Z^{\text{st}}(\gamma\tau,\gamma\bar\tau)) =  \frac12\(\frac{E\(\frac c2, \tau, \bar\tau\)}{\tau_2^\frac{c}2 |\eta(\tau)|^{2c}} + 2^c\left[\Big|\frac{\eta(\tau)}{\theta_2(\tau)}\Big|^c + \Big|\frac{\eta(\tau)}{\theta_3(\tau)}\Big|^c + \Big|\frac{\eta(\tau)}{\theta_4(\tau)}\Big|^c\right]\) 
\ee
which is precisely the average partition function (\ref{eq:avgorbz2}):
\be
\sum_{\gamma\in\Gamma_\infty \backslash SL(2,\mathbb Z)}(Z^{\text{st}}(\gamma\tau,\gamma\bar\tau))  = \langle Z_{T^c/\mathbb Z_2}(\tau,\bar\tau) \rangle_{\cM}\ .
\ee
We will give a bulk explanation of this sum in the next subsection.

\subsection{Bulk Chern-Simons Theory}\label{ss:CSZ2}

Let us now interpret this result from the bulk perspective. 
According to the standard dictionary of the AdS$_3$/CFT$_2$ correspondence, 
a massless free scalar field $X$ in two dimensions is dual to a pair of Chern-Simons 
gauge fields $A$ and $\tilde{A}$ with the boundary conditions, 
\begin{align}
&A_z =  \partial_z X, ~~ A_{\bar z}=0, \nn \\
&\tilde{A}_{\bar z} = \partial_{\bar z} X, ~~ \tilde{A}_z = 0,  
\label{GaugeBoundaryCondition}
\end{align}
in the gauge
where the component of each gauge connection normal to the boundary vanishes (see, for example, \cite{Kraus:2006wn}). 
This suggests that the ${\mathbb Z}_2$ symmetry, $X \rightarrow -X$, in two dimensions acts as
$A \rightarrow - A$ and $\tilde{A} \rightarrow - \tilde{A}$ in this gauge. 
This is a global symmetry in the Chern-Simons theory. Even though the bulk theory is meant to be gravitational,
the global symmetry is allowed here since the standard arguments for the absence of global symmetries 
assume weakly coupled Einstein gravity \cite{Banks:2010zn, Harlow:2018tng, Harlow:2018jwu}.

Orbifolding the free scalar $X$ by the ${\mathbb Z}_2$ symmetry should then be dual to gauging the ${\mathbb Z}_2$ global
symmetry of the Chern-Simons theory. Consider the Chern-Simons theory with gauge group 
$U(1)^{2c} \otimes \mathbb{Z}_2$ on the solid torus whose spatial cycle is contractible. 
To project onto ${\mathbb Z}_2$ invariant states, we insert $P = (1 + (-1)^{\sigma})/2$ in the partition function,
where $(-1)^{\sigma}$ generates the ${\mathbb Z}_2$ action on states on the spatial slice. 
The partition function with the insertion of $1$ is the same as the one computed by 
 \cite{Afkhami-Jeddi:2020ezh, Maloney:2020nni},
\be\label{Zpp}
Z_{++}=\frac1{|\eta(\tau)|^{2c}}\ .
\ee
On the other hand, the insertion of $(-1)^{\sigma}$ gives
\be
Z_{+-}=|q|^{-c/12}\prod_{n=1}^\infty\frac1{|1+q^n|^{2c}} = \left|\frac{\eta(\tau)}{\eta(2\tau)}\right|^{2c}\ .
\ee
To obtain this, we need to compute the functional determinants of twisted Laplacians as in appendix~C of \cite{Porrati:2019knx}. Twisting here means imposing a twisted boundary condition on the eigenfunctions of the Laplacians.
Projecting to $\sigma$-invariant states thus indeed gives
\be
\frac12 Z_{++}+\frac12 Z_{+-}= |\chi^{\text{vac}}|^2 + |\chi^1|^2\ .
\ee

There are other contributions we need to take into account. One of the hallmarks of discrete gauge symmetry, as opposed to
discrete global symmetry, is the presence of vortices \cite{Krauss:1988zc, Preskill:1990bm}. 
They are co-dimension two objects around which local fields are
acted by elements of the gauge group. In the ${\mathbb Z}_2$ case, the Chern-Simons gauge fields transform as
$A \rightarrow - A$ and $\tilde{A} \rightarrow - \tilde{A}$ around the non-trivial vortices. They correspond to
non-trivial ${\mathbb Z}_2$ bundles in the bulk and are  
 similar to the conical defects considered in the context of the pure Einstein gravity in three dimensions in \cite{Benjamin:2020mfz}. 
Given that such vortices exist in the bulk, we should take into account configurations with vortices running along
non-contractible cycles of the solid torus since they should be part of the ``summing over gauge field configurations in the bulk'' 
as they represent  non-trivial ${\mathbb Z}_2$ bundles.

 These vortices can end on the boundary, and the end points are identified with the 
twist fields of the orbifold theory. Since there are $2^c$ twist fields corresponding to the $2^c$ fixed points on 
$T^c/{\mathbb Z}_2$, we expect that there are $2^c$  different vortices.

To compute the partition function with a single vortex insertion, 
we apply the double cover trick of the orbifold theory to the bulk theory.
Consider a solid torus of the modulus $\tau$ and cut it open along the `$1$' direction to obtain a solid cylinder
of circumference 1 and length $\tau$. If there is a ${\mathbb Z}_2$ vortex running through the center, 
we map the solid cylinder to its double cover, a solid cylinder of circumference of 2, length $\tau$ and with no vortex, which is conformally equivalent to a cylinder of circumference 1 and length $\tau/2$. Therefore, the contribution to the partition function of $2c$ gauge fields
 is again (\ref{Zpp}), with $\tau$ replaced by $\tau/2$. This however is not quite what we are after: 
Each of the gauge fields $A(z)$ on the cover is completely arbitrary, and thus does not correspond to a configuration that satisfies $A \mapsto -A$ 
on the base. Instead it represents a configuration of two independent fields $A^{1,2}$ on the base that are exchanged around the vortex
as, $A^{1,2}\mapsto A^{2,1}$. From this we want to extract the contribution of $A =A^{1}-A^{2}$, which indeed has the correct boundary condition 
around the vortex. To do this, we can cancel out the contribution of $A^1+A^2$, which is simply the contribution without any vortex.
 We thus get
\be
Z_{-+} = \left| \frac{\eta(\tau)}{\eta(\tau/2)} \right|^{2c}=
 \left| q^{\frac{1}{48}}\prod_{n=1}^\infty \frac1{1-q^{n-\frac12}} \right|^{2c}
= \left| \frac{\eta(\tau)}{\theta_4(\tau)} \right|^c\ .
\ee
Inserting $(-1)^{\sigma}$ in the vortex sector works the same way as before and gives
\be
Z_{--} = \left| q^{\frac1{48}}\prod_{n=1}^\infty \frac1{1+q^{n-\frac12}} \right|^{2c}
= \left| \frac{\eta(\tau)}{\theta_3(\tau)} \right|^c\ .
\ee
Projecting to the invariant states in the vortex similar to the conical defects observed in \cite{Benjamin:2020mfz} sector, we find
\be
\frac12 Z_{-+}+\frac12 Z_{--}= |\chi^2|^2 + |\chi^3|^2 .
\ee
Since there are $2^c$ different vortices, the total Chern-Simons partition function for the solid torus, with and without vortices running in
the middle, is
\be
Z^{\text{st}} = |\chi^{\text{vac}}|^2 + |\chi^1|^2 + 2^c|\chi^2|^2 + 2^c |\chi^3|^2\ ,
\ee
which is indeed the same as $Z^{\text{st}}$ in eq. (\ref{Zin}).
Of course one can imagine more complicated configurations of vortices. However, due to the $\Z_2$ fusion rules of vortices, they can be reduced to either the vortex running through the solid torus once or no vortex at all.
A sum over all contractible cycles indeed gives the Poincar\'e series (\ref{Z2Poincare}).

\section{Averaging over $T^{2d}/\mathbb{Z}_N$}
\label{sec:tczn}
In this section we generalize our computations and consider $T^{2d}/\bZ_N$ orbifolds. The main difference to the $\bZ_2$ case is that not every $T^{2d}$ has a $\bZ_N$ symmetry. We will fix the action of the symmetry and then integrate only over deformations that are compatible with that symmetry. We therefore have to restrict ourselves to a sublocus of the moduli space that allows for such a symmetry; as we will explain below, we will 
average over the K\"ahler moduli space with fixed complex structure.

\subsection{Average over K\"ahler Moduli Space}\label{avgzn}

Let $\bZ_N$ be a symmetry acting on $\bR^c$, where we assume that $c=2d$. Let us define our $\mathbb Z_N$ action as the following. We assume that the symmetry has no fixed points other than the origin, and we choose complex basis vectors $\{ e^\mu_i, e^\mu_{\bar i}\}$ with $(e^\mu_i)^*=e^\mu_{\bar i}$ such that $\bZ_N$ acts as
\be
Z^i \mapsto e^{2\pi i/N} Z^i\ ,\qquad \bar Z^{\bar i} \mapsto e^{-2\pi i/N}  \bar Z^{\bar i}\ .
\label{eq:symmetryaction}
\ee
These vectors define a complex structure. We note that even though we are setting up our formalism for arbitrary $N$, only for the crystallographic cases $N=3,4,$ and 6 can we actually find lattices with the symmetry (\ref{eq:symmetryaction}). For other choices of action of $\mathbb{Z}_N$, however, other values of $N$ are possible \cite{BCK}.

Following \cite{Ferrara:1992gk}, we fix the complex structure by choosing complex vectors $\{ e^\mu_i, e^\mu_{\bar i}\}$ to express any real vector $X^\mu$ as 
\be
X^\mu = Z^i e^\mu_i+\bar Z^{\bar i} e^\mu_{\bar i}\ .
\ee 
This gives the metric and $B$-field in complex notation:
\be
G_{i\bar\jmath} = G_{\mu\nu}e^\mu_i e^\nu_{\bar\jmath}\ ,\qquad  B_{i\bar\jmath} = B_{\mu\nu}e^\mu_i e^\nu_{\bar\jmath}\ .
\ee
Since we want the CFT to be invariant under the rotation $Z^i \mapsto e^{2\pi i/N} Z^i, \bar Z^{\bar i} \mapsto e^{-2\pi i/N}  \bar Z^{\bar i}$, we need to fix the complex structure such that 
\be
G_{ij}=G_{\bar i\bar\jmath} = B_{ij} = B_{\bar i \bar\jmath}= 0\ .
\ee
$G_{i\bar\jmath}$ is a real $d\times d$ matrix and $G_{i\bar\jmath}=G_{\bar\jmath i}$. We have $G_{i\bar\jmath}G^{\bar\jmath k}=\delta^k_i$.
Moreover, we have
\be
B_{\bar\jmath i} = B_{\mu\nu}e^\mu_{\bar\jmath} e^\nu_{i}=- B_{i \bar\jmath}
\ee
and
\be
B_{\bar\jmath i}^* = B_{\mu\nu}e^\mu_{ j} e^\nu_{\bar i} = - B_{\bar i j}\ .
\ee
$G_{i\bar\jmath}$ and $B_{i\bar\jmath}$ are the coordinates for the $c^2/2$ real dimensional K\"ahler sublocus $\cM_K$ of $\cM$. The Siegel-Narain theta function is defined as
\be\label{thetacpx}
\Theta_m(\tau,\bar\tau) = \sum_{\vec n,\vec w}Q(\vec n,\vec m;m,\tau)
\ee
where
\be\label{latsumQ}
Q= \exp\left( -\frac{\pi\tau_2}{\alpha'}(2G^{r\bar s}v_r\bar v_{\bar s}+2 G_{r\bar s}w^r \bar w^{\bar s}) +2\pi i\tau_1 (n_rw^r +\bar n_{\bar r}\bar w^{\bar r)}
\right)
\ee
with 
\be\label{latsumQ_ii}
v_r = \alpha'n_r+B_{r\bar s} \bar w^{\bar s}\ , 
\qquad \bar v_{\bar r} = \alpha'\bar n_{\bar r}+B_{\bar r s} w^{ s}\ .
\ee
The upshot is that instead of averaging over the entire moduli space $\cM$, we keep the complex structure fixed and compatible with the $\bZ_N$ symmetry, and instead average only over the K\"ahler structure $\cM_K$ parametrized by $G_{i\bar\jmath}$ and $B_{i\bar\jmath}$.

We would like to compute the average of the partition function
\be
Z_{T^c}(\tau,\bar\tau) = \frac{\Theta_m(\tau,\bar\tau)}{|\eta(\tau)|^{2c}}
\label{Z_TD}
\ee
over $\cM_K$. As before, we first compute the average of $\Theta(m,\tau)$:
\be
\langle \Theta(m, \tau) \rangle = \int_{\cM_K} d\mu(m) \Theta(m,\tau)\ ,
\ee
where again we emphasize that we average only over $\mathcal{M}_K$, not the entire Narain moduli space.
To do this, we follow \cite{Maloney:2020nni}. We establish that $\Theta(m,\tau)$ satisfies the differential equation
\be\label{mainDE}
\Big(\Delta_{\cH} - c\tau_2\frac{\partial}{\partial\tau_2}-  \Delta_{\cM_K} \Big)\Theta(m,\tau)= 0\ .
\ee
Here $\Delta_{\cH}$ is the Laplacian on the upper half plane $\Delta_{\cH}=\tau_2^2(\partial^2_{\tau_2} + \partial^2_{\tau_1})$, and $\Delta_{\cM_K}$ is the Laplacian restricted to the K\"ahler sublocus $\cM_K$. In  appendix~\ref{app:Kahler} we derive its explicit form which is given by
\be
\Delta_{\cM_K} = - G_{m\bar p} G_{q\bar n} (\partial_{G_{m\bar n}} \partial_{G_{\bar pq}} + \partial_{B_{m\bar n}} \partial_{B_{\bar p q}}) \ .
\ee
Appendix~\ref{app:Kahler} then verifies that $\Theta(m,\tau)$ indeed satisfies (\ref{mainDE}). Averaging (\ref{mainDE}) over $\cM_K$ and integrating by parts establishes $(\Delta_{\cH} - c\tau_2\frac{\partial}{\partial\tau_2})\langle \Theta(m, \tau) \rangle=0$. Since $\langle \Theta(m, \tau) \rangle$ has modular weight $(c/2,c/2)$ and satisfies $\lim_{\tau_2\to\infty} \langle \Theta(m, \tau) \rangle=1$, the same argument as in \cite{Maloney:2020nni} establishes that
\be\label{Zaverage}
\langle Z_{T^c}(\tau, \bar\tau) \rangle_{\cM_K} = \frac{E\(\frac c2, \tau, \bar\tau\)}{\tau_2^\frac{c}2 |\eta(\tau)|^{2c}}\ .
\ee

\subsection{The Poincar\'e Series}
Let $g^i$ be the generators of $\mathbb Z_N$. The twisted sectors are defined by elements $g^i$ which act as rotations on $T^{c}$. We can insert an element $g^j$, which gives the twisted twining partition function $Z_{(i,j)}$. The partition function of the orbifold CFT is then given by
\be\label{Zorb}
Z_{T^c/\bZ_N}(\tau,\bar\tau)= \frac1{N} \Big( Z_{(0,0)} + \sum_{(i,j)\neq(0,0)}Z_{(i,j)}\Big)\ .
\ee
Here the first term and the sum are separately modular invariant. In fact, $Z_{0,0}$ is simply the unorbifolded partition function
\be
Z_{0,0}(\tau,\bar\tau)= Z_{T^c}(\tau,\bar\tau)\ .
\ee
Since the action of $\mathbb Z_N$ on the lattice $\Lambda\subset \mathbb R^c$ has no fixed points other than the origin, the twined untwisted sectors with $j\neq 0$ do not contain any winding and momentum modes and are therefore independent of the moduli. Moreover, since all twisted sectors can be obtained as modular transformations of the twined untwisted sector, it follows that they are also independent of the moduli. The upshot is thus that in (\ref{Zorb}), only the first term depends on the moduli. Using (\ref{Zaverage}) it follows that
\be
\langle Z_{T^c/\bZ_N}(\tau,\bar\tau)\rangle_{\cM_K}=\frac1N\left(\frac{E\(\frac c2, \tau, \bar\tau\)}{\tau_2^\frac{c}2 |\eta(\tau)|^{2c}}
+\sum_{(i,j)\neq(0,0)}Z_{(i,j)}(\tau,\bar\tau)\right)\ .
\label{eq:abcdefg}
\ee
To compute the Poincar\'e series, we write
\be
Z_{(0,0)}= Z_{T^c}^{\text{vac}} + \sum_{h,\bar h}Z_{h,\bar h}\ ,
\ee
where $Z_{T^c}^{\text{vac}}$ is the contribution of all vacuum descendants of the unorbifolded theory, which is therefore moduli independent. The only moduli dependence is thus contained in the sum over $h,\bar h$. The contribution of the moduli independent terms is therefore given by
\be\label{Zshort}
Z^{\text{st}}= \frac1{N} \bigg( Z^{\text{vac}}_{T^c} + \sum_{(i,j)\neq(0,0)}Z_{(i,j)}\bigg)\ .
\ee
(We again follow the notation of Section \ref{sec:tcz2} and call this quantity $Z^{\text{st}}$, in anticipation that this will be a bulk quantity computed on the solid torus.) 
Our proposal is then that the Poincar\'e series of $Z^{\text{st}}$ is equal  to the average over the K\"ahler moduli space $\cM_K$,
\be\label{ZNprop}
 \sum_{\gamma \in \Gamma_\infty \backslash SL(2,\mathbb Z)}Z^{\text{st}}(\gamma\tau,\gamma\bar\tau)
= \langle Z_{T^c/\bZ_N}(\tau,\bar\tau)\rangle_{\cM_K}\ .
\ee
This follows immediately from what we have said so far: The second term in (\ref{Zshort}) is modular invariant and (with the Poincar\'e series of $1$ regularized to be 1) remains unchanged under the Poincar\'e series. On the other hand, $Z_{T^c}^{\text{vac}}$ is the vacuum sector of the original unorbifolded torus partition function, whose Poincar\'e series is the first term in (\ref{eq:abcdefg}). This establishes (\ref{ZNprop}).

\subsection{Bulk Chern-Simons Theory}
Let us briefly sketch how we obtain the Poincar\'e sum of $Z^{\text{st}}$ from a bulk computation. The story here is very similar to what we described in section~\ref{ss:CSZ2}. Rather than only $(-1)^\sigma$ we can now insert operators $g^i$ in the path integral, compute such twisted Laplacian determinants as in \cite{Porrati:2019knx}, and use the results to project to invariant states.

We again introduce vortices to extend the boundary twist fields into the bulk. Now there are $N$ types of such vortices which satisfy the fusion rule
\be
[i]+[j] = [i+j]\ .
\ee
The term $Z_{i,j}$ corresponds to the Chern-Simons path integral with vortex $[i]$, or alternatively, $i$ copies of vortices $[1]$ and operator $g^j$ inserted. We conjecture that these Chern-Simons path integrals will indeed give the corresponding terms in $Z^{\text{st}}$ in (\ref{Zshort}). To establish this, we expect that a similar argument as in section~\ref{ss:CSZ2} will apply here.

\section{Averaging correlation functions of $\mathbb{Z}_2$ orbifolds}\label{s:corrfcts}

In the previous sections, we have 
shown that ensemble averages of the partition functions of
two-dimensional CFTs with orbifold target spaces have holographic interpretations in terms of exotic bulk gravity theories
in three dimensions, generalizing the results of \cite{Afkhami-Jeddi:2020ezh, Maloney:2020nni}
for the torus target space. A new feature is the ${\mathbb Z}_N$ gauge symmetry in the bulk, which generates
a sum over  ${\mathbb Z}_N$-bundles on the solid torus. 
 One can now ask whether the correspondence can be
extended to other observables in CFT, such as correlations functions. 

For the torus target space of \cite{Afkhami-Jeddi:2020ezh, Maloney:2020nni}, 
the correspondence is trivial for correlation functions. 
Correlation functions of the $U(1)_L^c \times U(1)_R^c$ currents do not depend on the Narain moduli of $T^c$,
and it is trivial to average them. They obviously correspond to boundary-boundary correlation
functions of the $U(1)^c \times U(1)^c$ Chern-Simons theory in three dimensions.
One may try to find bulk duals of correlation functions
involving non-zero winding and momentum
numbers, which depend non-trivially on the moduli.
However, since the bulk theory is a pure Chern-Simons theory coupled to topological gravity,
there are no states which carry
 non-zero charges with respect to the $U(1)^c \times U(1)^c$
symmetry.\footnote{More precisely the partition function on the solid torus does not contain any charged states. See \cite{Datta:2021ftn} for further discussions.} 
 One may still consider correlation functions of Wilson-lines in the bulk ending
on the boundary at the insertion points of charged operators, but they do not 
reproduce averages of the CFT correlation functions with non-zero winding and momentum
numbers. In fact, averaging these CFT correlation functions over the Narain moduli space gives 
transcendental functions of insertion points of these operators. On the other hand,
the correlation functions of Wilson-lines in the bulk 
are elementary functions of the end-points of the Wilson lines. Since there are only a finite number of
ways to connect CFT operators by Wilson lines, the correlation function remains an elementary function
even after we sum over the possible configurations. Thus, the correspondence fails for correlation functions
with non-zero winding and momentum numbers. 

In this section, we will find a non-trivial duality for correlation functions in the orbifold theories. 
The partition function computation in the previous section suggests that the
bulk dual to the $\mathbb Z_N$ orbifold of $T^c$ is a $U(1)^c \times U(1)^c$ Chern-Simons theory 
with its ${\mathbb Z}_N$ global symmetry gauged, coupled to topological gravity which sums over three-dimensional topologies
allowing hyperbolic metrics. In this case, we can identify holographic duals of correlation functions of twist operators in the CFT. 
We show that the averages of the twist operator correlation functions over the orbifold moduli space
are equal to sums of correlation functions of ${\mathbb Z}_N$ vortices ending on the insertion points of the twist operators, and
that the sums are 
over configurations of these vortices that are compatible with hyperbolic structure in the bulk. 
In this section, we focus on the $N=2$ case. First, we will 
compute the averages of CFT correlation functions of twist operators in section \ref{corr_Z2_cft}. We 
will then interpret the results from the bulk perspective in section \ref{corr_Z2_bulk}.  Averages of correlation functions of $\bZ_N$ orbifold theories with $N>2$ will be discussed in section \ref{sec:ZNCor}. The idea of interpreting correlation functions in CFTs as a sum over bulk vortex configurations was first considered in \cite{Maloney:2016kee}. In this section we will show that the twist operator four-point functions in the averaged Narain orbifold theory provide a concrete realization of what was anticipated in \cite{Maloney:2016kee}. 

\subsection{Ensemble Average of CFT Correlation Functions}\label{corr_Z2_cft}

On $T^c/\bZ_2$, there are $2^c$ fixed points, which can be labeled by vectors $\vec\epsilon \in \bZ_2^c$. Consider a correlation function of four twist fields 
$\sigma_{\vec\epsilon_i}$ ($i = 0, 1, 2, 3$) on $S^2$. By the charge conservation,  $\sum_{i=0}^3 \vec\epsilon_i =0$. Moreover, we can always choose the origin of $T^c$ so that one of the vectors is zero. Thus,
the most general four point function is given by 
\be
G_{\vec\epsilon_0,\vec\epsilon_1}(x)=\langle \sigma_{\vec\epsilon_0}(0)\sigma_{\vec\epsilon_1}(1)
\sigma_{\vec\epsilon_1+\vec\epsilon_0}(x)\sigma_0(\infty) \rangle\ .
\label{eq:mostgenz2cor}
\ee
The method to evaluate this by going to the double cover of  $S^2$ was developed 
in \cite{Hamidi:1986vh, Dixon:1986qv}.

\subsubsection{$W_{\bf\epsilon_0,\epsilon_1}(\tau)$}

The correlation function (\ref{eq:mostgenz2cor}) is given by \cite{Dixon:1986qv}
\be
G_{\vec\epsilon_0,\vec\epsilon_1}(x)=
2^{-2c/3}|x(1-x)|^{-c/12}\frac1{|\eta(\tau)|^{2c}}\Theta_{\bf\epsilon_0,\epsilon_1}(m,\tau)\ ,
\ee
where
\be\label{Theta4pt}
\Theta_{\bf\epsilon_0,\epsilon_1}(m,\tau):=\sum_{\substack{\bold w\in\bZ^c \\ \bold n\in2\bZ^c+\bf\epsilon_0}}
(-1)^{\bold w\bf\epsilon_1}e^{-\frac{2\pi\tau_2}{\alpha'}(G^{r\bar s}v_r\bar v_{\bar s}+G_{r\bar s}w^r\bar w^{\bar s})+
	2\pi i\tau_1(n_rw^r+\bar n_{\bar r}\bar w^{\bar r})}\ ,
\ee
and the modulus of the covering torus $\tau$ is related to the cross ratio $x$ by the modular $\lambda$ function
\be
1- x(\tau)=\lambda(\tau) = \frac{\theta_2^4(\tau)}{\theta_3^4(\tau)}\ .
\label{LambdaFunction}
\ee
The Siegel-Narain theta function $\Theta_{\bf\epsilon_0,\epsilon_1}(m,\tau)$ satisfies the same differential equation as in \cite{Maloney:2020nni} (see section \ref{sec:ReviewNarain}), namely
\be\label{mainDE4pt}
\Big(\Delta_{\cH} - c\tau_2\frac{\partial}{\partial\tau_2}-  \Delta_{\cM} \Big)\Theta_{\bf\epsilon_0,\epsilon_1}(m,\tau)=0\ .
\ee
We define the average of $\Theta_{\bf\epsilon_0,\epsilon_1}$ over $\cM$,
\be\label{F01}
W_{\bf\epsilon_0,\epsilon_1}(\tau) := \tau_2^{c/2} \int_{\cM} d\mu(m) \Theta_{\bf\epsilon_0,\epsilon_1}(m,\tau)\ .
\ee
(In (\ref{F01}), we introduced an extra factor of $\tau_2^{c/2}$ for convenient to make $W_{\bf\epsilon_0,\epsilon_1}(\tau)$ modular invariant under the congruence subgroup $\Gamma(2)$.)
Assuming that the integral over the moduli space converges, (\ref{mainDE4pt}) tells us that $W_{\bf\epsilon_0,\epsilon_1}$ satisfies
\be
\left(\Delta_{\cH} - c\tau_2\frac{\partial}{\partial\tau_2}\right)\(\tau_2^{-c/2}W_{\bf\epsilon_0,\epsilon_1}(\tau)\)=0\ .
\ee
Under the finite group $SL(2,\bZ)/\Gamma(2)$, $W_{\bf\epsilon_0,\epsilon_1}(\tau)$ transforms as a vector valued function \cite{Keller:2019yrr}:
\begin{align}
&\gamma:&&
\begin{pmatrix} 1&0\\0&1\end{pmatrix}&&
\begin{pmatrix} 0&-1\\1&0\end{pmatrix}&&
\begin{pmatrix} 1&0\\-1&1\end{pmatrix}&&
\begin{pmatrix} 1&-1\\0&1\end{pmatrix}&&
\begin{pmatrix} 0&1\\-1&1\end{pmatrix}&&
\begin{pmatrix} 1&-1\\1&0\end{pmatrix}\\
&W: &&\;\;\; W_{\e_0,\e_1}&&\;\;\;\;W_{\e_1,\e_0}&&\;\;W_{\e_0+\e_1,\e_1}&&\;\;W_{\e_0,\e_0+\e_1}
&&\;\;\;W_{\e_1,\e_0+\e_1}&&\;\;W_{\e_0+\e_1,\e_0}\label{Ztrafo}
\end{align}
$\Gamma(2)$ has the cusps $\{0,1,i\infty\}$. Physically, it is clear that $W_{\bf\epsilon_0,\epsilon_1}(\tau)$ is finite on ${\cal H}/\Gamma(2)$ away from these cusps, where $\cal H$ is the upper half plane. 

\subsubsection{Eigenfunctions of the Laplacian}
The function $W_{\bf\epsilon_0,\epsilon_1}$ introduced in eq. (\ref{F01}) is defined on ${\cal H}/\Gamma(2)$ and satisfies the differential equation
\be\label{LaplaceEigen}
\left(\Delta_{\cH} +\frac c2 (\frac c2-1)\right)W_{\bf\epsilon_0,\epsilon_1}(\tau)=0\ .
\ee
That is, $W_{\bf\epsilon_0,\epsilon_1}$ is an eigenfunction of the Laplacian of eigenvalue $-\frac c2 (\frac c2-1)$. In particular, if $c>2$, it is an eigenfunction with negative eigenvalue.

We next describe the space of such eigenfunctions $W_{\bf\epsilon_0,\epsilon_1}$ defined on ${\cal H}/\Gamma(2)$.
First, let us give three examples of such eigenfunctions. There is the Eisenstein series given by\footnote{We hope that the two uses of $c$ in (\ref{EiInf}) as both the central charge and the lower-left entry of an element of $\Gamma(2)$ will not confuse the reader.}
\be\label{EiInf}
E_{i\infty}^{\Gamma(2)}(c/2,\tau,\bar\tau)= \sum_{\gamma\in\Gamma_\infty\backslash\Gamma(2)} \text{Im} (\gamma\tau)^{c/2}
= \sum_{\gamma\in\Gamma_\infty\backslash\Gamma(2)} \frac{\tau_2^{c/2}}{|c\tau+d|^c} \ .
\ee
There are 6 images of this under $SL(2,\bZ)/\Gamma(2)$, but because (\ref{EiInf}) is invariant under $T$ modular transformation, there are only two new images:
\bea
E_0^{\Gamma(2)}(c/2,\tau,\bar\tau)&:=& E_{i\infty}^{\Gamma(2)}(c/2,-1/\tau,-1/\bar\tau)\ ,\\
E_1^{\Gamma(2)}(c/2,\tau,\bar\tau)&:=& E_{i\infty}^{\Gamma(2)}(c/2,-1/(\tau-1),-1/(\bar\tau-1))\ .
\eea
These functions are regular on ${\cal H}/\Gamma(2)$, but they diverge at the cusps.
More precisely, they have the following behavior:
\be\label{EiInfcusp}
E_{i\infty}^{\Gamma(2)}  \xrightarrow[\tau\to i\infty]{}\tau_2^{c/2}.
\ee
We can also expand $E_0^{\Gamma(2)}(\tau)$ at the other cusps (see eq. (C.22) of \cite{Benjamin:2020zbs}) to get: 
\be\label{Ei01cusp}
E_{0}^{\Gamma(2)} \xrightarrow[\tau\to i\infty]{} 0\ , \qquad E_{1}^{\Gamma(2)} \xrightarrow[\tau\to i\infty]{} 0\ .
\ee
The reason for this is that in the Eisenstein series, the only element that does not go to 0 as $\tau\to i\infty$ is the identity element (and its $T$ transforms). Since that element is not in the sum $E_{0}^{\Gamma(2)}$ and $E_{1}^{\Gamma(2)}$, they behave as (\ref{Ei01cusp}). Note that from (\ref{EiInfcusp}) and (\ref{Ei01cusp}) we can immediately read off the behavior at the cusps. Namely, $E_s^{\Gamma(2)}$ diverges at the cusp $s$, and vanishes at the two other cusps. 

This divergence immediately implies that the Eisenstein series are not square integrable. That is not surprising, since they are eigenfunctions of the Laplacian with negative eigenvalue. We know however that on $\cL^2$, the Laplacian is positive definite.

We now claim that $W_{\bf\epsilon_0,\epsilon_1}$ can be written as a linear combination of the three Eisenstein series,
\be \label{WABC}
W_{\bf\epsilon_0,\epsilon_1}(\tau)= A E_{i\infty}^{\Gamma(2)}(c/2,\tau,\bar\tau)
+ B E_0^{\Gamma(2)}(c/2,\tau,\bar\tau)
+ C E_1^{\Gamma(2)}(c/2,\tau,\bar\tau)\ ,
\ee
where the three coefficients $A,B,C$ depend on $\epsilon_0$ and $\epsilon_1$, and we fix them by matching at each cusp $0,1,i\infty$ the behavior of $W_{\bf\epsilon_0,\epsilon_1}(\tau)$ to the behavior of the $E^{\Gamma(2)}$.
To establish (\ref{WABC}), let us take the difference of the left- and right-hand side. The resulting function is clearly still an eigenfunction of the Laplacian with negative eigenvalue. We claim that it is square integrable, and hence vanishes.

To see this, we first note that all the functions appearing in (\ref{WABC}) are regular away from the cusps. We thus only need to establish that their difference is square integrable around $i\infty, 0$ and 1. Let us first discuss the behavior at $i\infty$. We know that $W_{\bf\epsilon_0,\epsilon_1}(\tau) \sim \tau_2^{c/2}$ as $\tau\to i\infty$. Since by (\ref{Ei01cusp}) we know that the two other Eisenstein series remain regular, we simply fix $A$ such that the leading divergence cancels the divergence (\ref{EiInfcusp}). It then follows that the difference grows more slowly than $\tau_2^{c/2}$.
However, (\ref{LaplaceEigen}) implies that in the limit $\tau\to i\infty$ any eigenfunction behaves as
\be
W_{\bf\epsilon_0,\epsilon_1}(\tau) \sim a\tau_2^{c/2} + b\tau_2^{1-c/2}\ .
\ee
Since the difference grows slower than $\tau_2^{c/2}$, it must grow like $\tau_2^{1-c/2}$, and is thus square integrable around the cusp $i\infty$.

To deal with the cusps 0 and 1, we simply map them to $i\infty$ using an appropriate element of $SL(2,\bZ)$. Since both the integration measure and $\Delta_{\cal H}$ are invariant under such transformations, the only thing that changes is that the three Eisenstein series are permuted. To ensure integrability around the other cusps, it is thus necessary to fix $B$ and $C$ to match the leading divergence at those cusps coming from $E_0^{\Gamma(2)}(c/2,\tau,\bar\tau)$ and $E_1^{\Gamma(2)}(c/2,\tau,\bar\tau)$. The difference is then square integrable around all cusps, and therefore necessarily vanishes, establishing (\ref{WABC}).

\subsubsection{Fixing the cusps}

To fix the coefficients $A$, $B$, and $C$ in eq. (\ref{WABC}), we compute the behavior of $W_{\bf\epsilon_0,\epsilon_1}$ near the cusps. To do this, let us consider eq. (\ref{Theta4pt}). In the limit $\tau\to i\infty$, we only need to evaluate the leading term. This is the term with ${\bf w}=0$. If ${\bf \epsilon_0}=0$, then there is a term with ${\bf n}=0$, so that $\Theta \to 1$. On the other hand if ${\bf \epsilon_0}\neq0$, then every term has a strictly positive power of $q$, so that $\Theta \to 0$. After exchanging the limit and the integral in eq. (\ref{F01}), and using the fact that the integration measure $d\mu(m)$ is normalized to one, we obtain
\be
\lim_{\tau\to i\infty} \tau_2^{-c/2} W_{\bf\epsilon_0,\epsilon_1}(\tau) = \delta_{\bf\epsilon_0,0}\ .
\ee
To obtain the behavior at the other cusps, we use (\ref{Ztrafo}) to map $W_{\bf\epsilon_0,\epsilon_1}(\tau)$ to other cusps. This gives
\be
\lim_{\tau\to 0} \(\text{Im}(-1/\tau)\)^{-c/2} W_{\bf\epsilon_0,\epsilon_1}(\tau) = 
\lim_{\tau\to i\infty} \tau_2^{-c/2} W_{\bf\epsilon_1,\epsilon_0}(\tau)
=\delta_{\bf\epsilon_1,0}\ .
\ee
and
\be
\lim_{\tau\to 1} \(\text{Im}(-\frac1{\tau-1})\)^{-c/2} W_{\bf\epsilon_0,\epsilon_1}(\tau) = 
\lim_{\tau\to i\infty} \tau_2^{-c/2} W_{\bf\epsilon_0+\epsilon_1,\epsilon_1}(\tau)
=\delta_{\bf\epsilon_0+\epsilon_1,0}\ .
\ee
In total, we thus obtain
\be
W_{\bf\epsilon_0,\epsilon_1}(\tau)= \delta_{\bf\epsilon_0,0}E_{i\infty}^{\Gamma(2)}(c/2,\tau,\bar\tau)
+ \delta_{\bf\epsilon_1,0}E_0^{\Gamma(2)}(c/2,\tau,\bar\tau)
+ \delta_{\bf\epsilon_1,\bf\epsilon_0}E_1^{\Gamma(2)}(c/2,\tau,\bar\tau)\ .
\label{TwstCorrAverage}
\ee
Note that this implies a stronger version of $\bZ_2^c$-charge conservation: there is a Wick-type contraction, but only between fields with matching charges.

\subsection{Bulk Interpretation}\label{corr_Z2_bulk}

In this subsection, we present a bulk interpretation of the CFT computation (\ref{TwstCorrAverage}). We shall discuss 
the ${\mathbb Z}_N$ orbifold with $N=2$ and will later generalize the results to $N > 2$ in section \ref{sec:ZNCor}. 

\subsubsection{Vortices for Discrete Gauge Symmetry}

If there is a ${\mathbb Z}_2$ twist operator at $z= z_0$ on the boundary, the CFT variables $X^\mu$ change their signs as they go around this point. In the bulk, this sign change is described by a ${\mathbb Z}_2$ gauge vortex emanating from the point $z_0$ on the boundary. In the gauge where the component of each gauge connection along the vortex vanishes, the gauge connections change their signs as they go around the vortex.

Since  ${\mathbb Z}_2$ vortices cannot split or join, the vortex emanating 
from $z_0$ should end somewhere on the boundary, say at $z=z_1$, where there must be another twist operator. 
Each twist operator is associated to a ${\mathbb Z}_2$ fixed point on $T^c$. 
It turns out that the fixed points associated to the two end-points of the vortex must be the same. To see this,
we note that the vortex cannot make a non-trivial knot in the bulk for a reason to be explained later. Therefore,
the vortex configuration must be homotopic to a curve connecting $z_0$ to $z_1$ on the boundary. Since the gauge connections of the
Chern-Simons theory must be flat, the integral of $A$ from  $z_0$ to $z_1$ on the boundary must be the same
as the integral of $A$ along the vortex, which should vanish by the gauge condition. 
Since the gauge connection is related to $X$ on the boundary by (\ref{GaugeBoundaryCondition}), 
the integral of $\partial_z X dz +   \partial_{\bar z} X d {\bar z}$ along the curve connecting
$z_0$ to $z_1$ should also vanish, namely, $X^\mu(z_0) = X^\mu(z_1)$. Thus, the  ${\mathbb Z}_2$  fixed
points associated to the two end-points of the vortex must coincide. This explains the selection rule in  (\ref{TwstCorrAverage})
imposed by the three Kronecker deltas, $ \delta_{\bf\epsilon_0,0}$, $\delta_{\bf\epsilon_1,0}$,
and $\delta_{\bf\epsilon_1,\bf\epsilon_0}$.

Let us focus on the four point function of twist operators discussed in the previous section. The four points are pairwise connected by two vortices in the bulk, and the two end-points of each vortex must be associated to the same ${\mathbb Z}_2$
fixed point, as explained in the above paragraph. The first term on the right-hand side of (\ref{TwstCorrAverage}), for example,
should correspond to a configuration where one vortex connects twist fields at $z=0$ and $\infty$
and another connects those at $z=x$ and $1$. 

There are infinitely many ways to connect $z=0$ and $\infty$ by a vortex and to connect $z=x$ and $1$ by another vortex.
The vortices can be linked in the bulk in topologically non-trivial manners, and each of them can knot with itself. 
As we are going to show in the next subsection,  there are topological restrictions on configurations of the vortices. 
With these restrictions, the Eisenstein series (\ref{EiInf}) computed by the ensemble average of the CFT
correlation function can be interpreted as a sum over restricted configurations of these vortices. 

\subsubsection{Rational Tangles}

Before deriving the topological restrictions on configurations of the vortices from the bulk perspective, it would be helpful to 
discuss the geometric meaning of the sum over ${\Gamma_\infty} \backslash \Gamma(2)$ in the Eisenstein series (\ref{EiInf})
from the CFT perspective.

The congruence subgroup $\Gamma(2)$ of the modular group acts on $\tau$ as the fractional linear transformation, 
and $\tau$ is related to 
the cross ratio $x$ of the four insertion points of the twist operators by $x = 1-\lambda(\tau)$ as in eq. (\ref{LambdaFunction}).
The inverse of this relation is 
\be
\tau(x) = i \frac{_2F_1(1/2, 1/2, 1:x)}{_2F_1(1/2, 1/2, 1: 1-x)}\ . 
\ee
Since 
the Frobenius method gives
\be
_2F_1(1/2, 1/2, 1: 1-x) = - \frac{1}{\pi} \log x \cdot ~_2F_1(1/2, 1/2, 1: x) + \cdots\ , 
\ee
where $(\cdots)$ represents a function that is single-valued around $z=0$, we find the monodromy relation
\be
_2F_1(1/2, 1/2, 1: 1 - e^{2 \pi i} x) = - 2i ~_2F_1(1/2, 1/2, 1: x)  + ~_2F_1(1/2, 1/2, 1: 1-x)\ .
\ee
Therefore, 
\be\label{STS}
\tau(e^{2 \pi i} x) = \frac{\tau(x)}{-2\tau(x) + 1}\ .
\ee
Similarly,
\be\label{TT}
\tau(e^{2 \pi i} (x-1) + 1) = \tau(x) - 2\ .
\ee
Since $\tau \rightarrow - 2 \tau + 1$ and $\tau \rightarrow \tau - 2$ generate $\Gamma(2)$, we
can identify $\Gamma(2)$ as the monodromy group of $x$ going around $0, 1,$ and $\infty$. Since both 
$\Gamma(2)$ and the monodromy group are free groups with two generators, this gives an isomorphism between the two groups. 

\begin{figure}
\centering
\subfloat[]{
\centering
\begin{tikzpicture}
\draw[thick, black] (2,3) -- (0,2) -- (0,-1.5) -- (2,-0.5) -- (2,-0.2);
\draw[thick, black] (2,-0.05) -- (2,0.5);
\draw[thick, black] (2,0.7) -- (2,1.3);
\draw[thick, black] (2,1.5) -- (2,2.05);
\draw[thick, black] (2,2.2) -- (2,3);
\draw[thick, red] (1.25,2.05) .. controls (4.5,2.5) and (4.5,1.6) .. (1.25,1.3);
\draw[thick, blue] (1.25,0.5) .. controls (4.5,1.1) and (4.5,-0.1) .. (1.25,-0.2);
\draw[thick, red, fill=red] (1.25,2.05) circle (0.75mm);
\draw[thick, red, fill=red] (1.25,1.3) circle (0.75mm);
\draw[thick, blue, fill=blue] (1.25,0.5) circle (0.75mm);
\draw[thick, blue, fill=blue] (1.25,-0.2) circle (0.75mm);
\node at (0.8,2.05) {0} node at (0.8,1.3) {$\infty$} node at (0.8,0.5) {$x$} node at (0.8,-0.2) {1};
\end{tikzpicture}
\label{rtan_i}
}
\subfloat[]{
\begin{tikzpicture}
\draw[->, double distance=2.5pt] (-4,1) -- (-1.75,1);
\node at (-2.9,0.5) {\large{monodromy}} node at (-2.6, 0) {\large{transformations}};
\draw[thick, black] (2,3) -- (0,2) -- (0,-1.5) -- (2,-0.5) -- (2,-0.2);
\draw[thick, black] (2,-0.05) -- (2,0.5);
\draw[thick, black] (2,0.7) -- (2,1.3);
\draw[thick, black] (2,1.5) -- (2,2.05);
\draw[thick, black] (2,2.2) -- (2,3);
\draw[thick, red] (1.25,2.05) .. controls (4.0,2.4) and (4,2.2) .. (3.9,1.8);
\draw[thick, red]  (3.7,1.6) .. controls (2.3,1.4) .. (1.25,1.3);
\draw[thick, blue] (1.25,-0.2) .. controls (4,-0.3) and (4.8,2) .. (3.4,1.7);
\draw[thick, blue] (1.25,0.5) .. controls (2.5,0.6) and (3,1) .. (3.3,1.4);
\draw[thick, red, fill=red] (1.25,2.05) circle (0.75mm);
\draw[thick, red, fill=red] (1.25,1.3) circle (0.75mm);
\draw[thick, blue, fill=blue] (1.25,0.5) circle (0.75mm);
\draw[thick, blue, fill=blue] (1.25,-0.2) circle (0.75mm);
\node at (0.8,2.05) {0} node at (0.8,1.3) {$\infty$} node at (0.8,0.5) {$x$} node at (0.8,-0.2) {1};
\end{tikzpicture}
\label{rtan_ii}}
\caption{Two rational tangles related by a monodromy transformation.}
\label{fig_rtangle}
\end{figure}
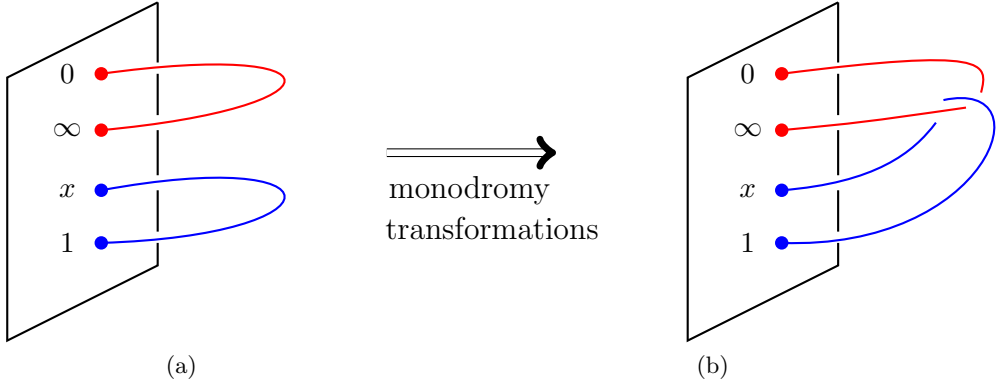

Each term in (\ref{EiInf}) is identified as a configuration of the two vortices in the bulk as follows. 
Consider a configuration of the vortices as in Fig. \ref{rtan_i}. If we move $x$ around $0$ and bring it back to the same point, 
a new vortex configuration depicted
in Fig. \ref{rtan_ii} is generated. In this way, the monodromy group of $x$ generates a set of vortex configurations. 
We will use the concept of \emph{rational tangles} to describe them. 
In knot theory, a 2-tangle is a proper embedding of the disjoint union of two arcs into a 3-ball
such that the endpoints of the arcs land on four marked points on the ball's boundary. A tangle is called rational 
if it is homeomorphic to the trivial one as in Fig. \ref{rtan_i}. It is known that rational 2-tangles are generated by 
acting the monodromy group on endpoints of the trivial tangle. Since the monodromy of $x$ around $1$ acts
trivially on the configuration depicted in Fig. \ref{rtan_i}, one can identify ${\Gamma_\infty} \backslash \Gamma(2)$ as
the set of rational 2-tangles pair-wisely connecting $0$ and $\infty$ and connecting $x$ and $1$ on the boundary.

 \begin{figure}
  \centering
\captionsetup{width=1\linewidth}
    \includegraphics[width=1\textwidth]{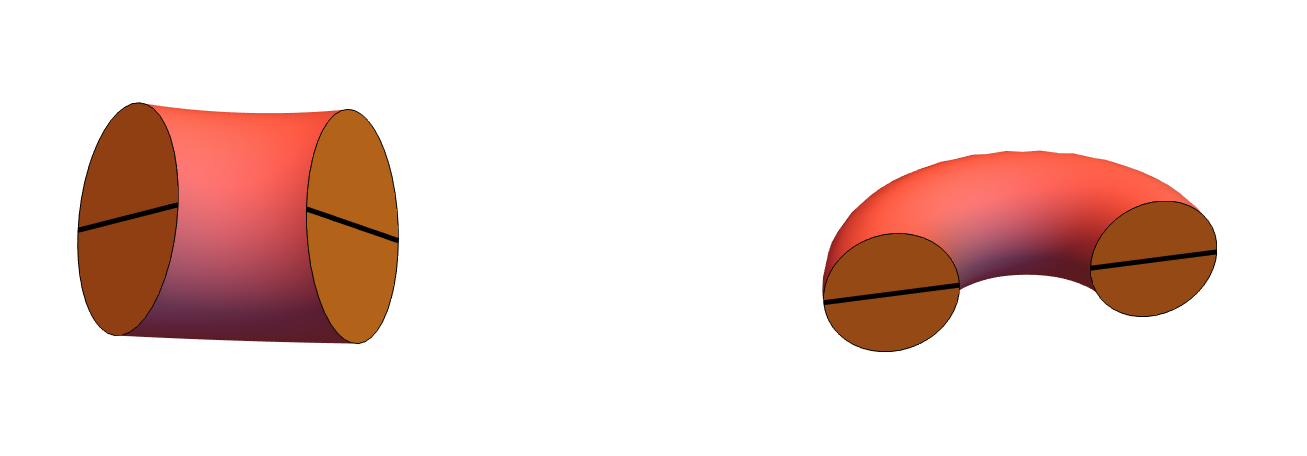}
      \caption{Consider a minimal surface bounded
by each vortex and the boundary of the 3 ball, and cut open the 3-ball along the two minimal surfaces as shown in the diagram. The pink region is the boundary, the black lines are the two-tangles, and the orange surfaces the branch sheets (or in the language of \cite{Thurston}, the orange surfaces are the doorways to Narnia). Gluing two copies of 3-balls across these minimal surfaces gives a solid torus of modulus $\tau$.}
\label{3ball}
\end{figure}

\begin{figure}%
    \centering
    \subfloat[][\centering]{\includegraphics[width=100pt]{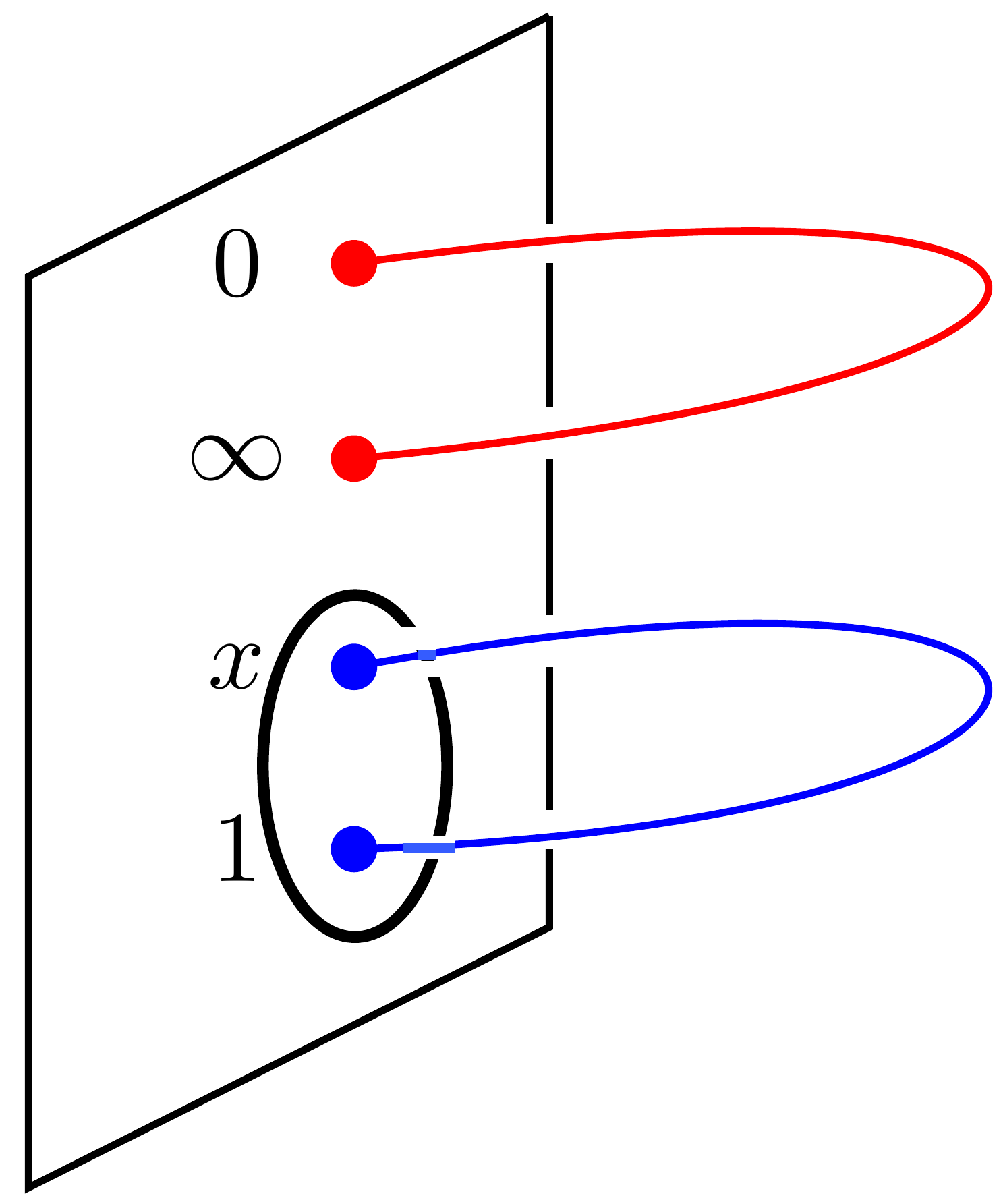} \label{fig:cyclea} }%
    \hspace{60pt}
    \subfloat[][\centering]{\includegraphics[width=100pt]{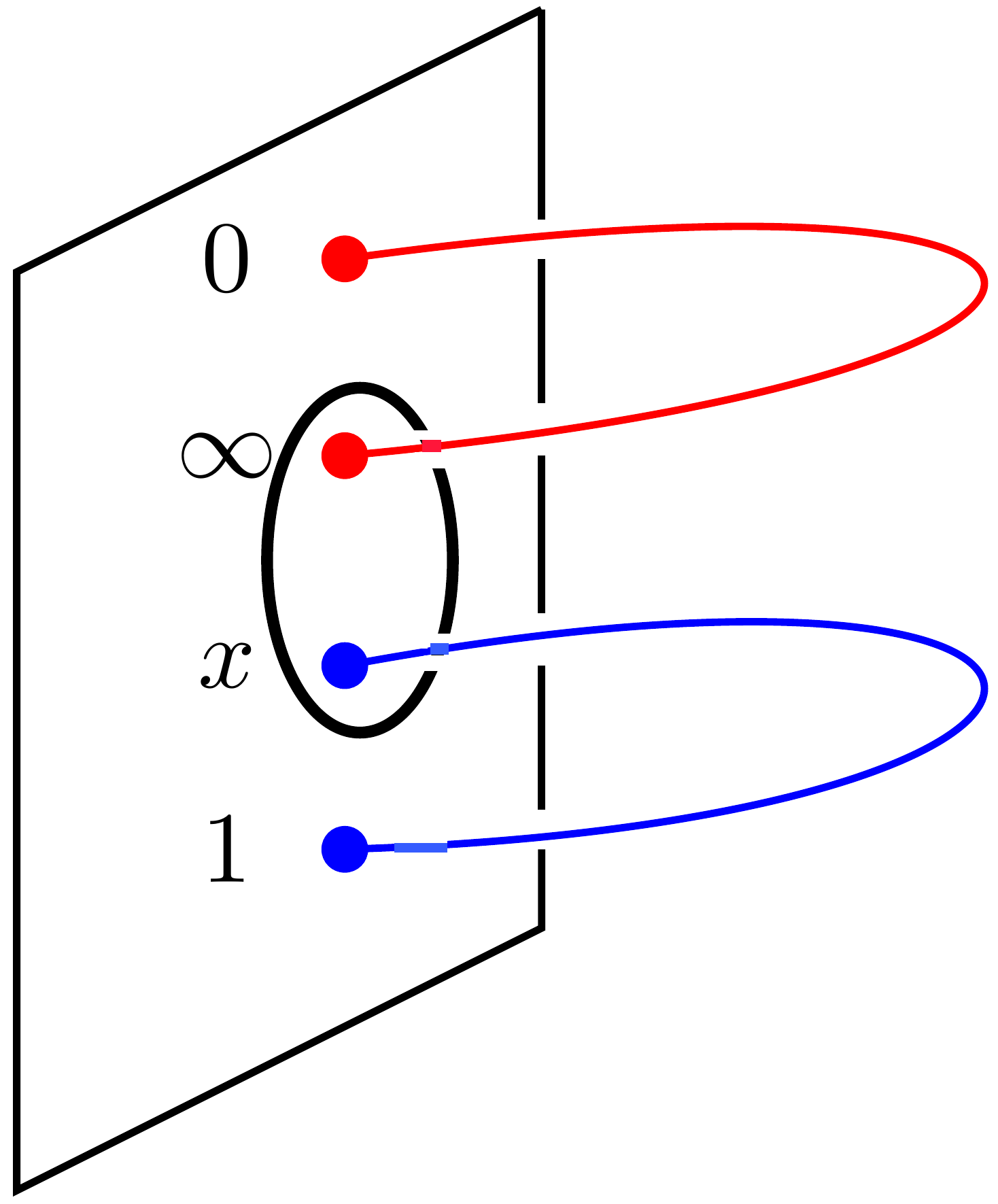}  \label{fig:cycleb} }%
    \caption{(a) The contractible cycle. (b) The non-contractible cycle}%
    \label{fig:cycles}%
\end{figure}

Let us discuss the bulk Chern-Simons computation for the vortex configuration depicted in Fig. \ref{rtan_i}. 
Since the gauge connections change their signs around each vortex, we can compute the Chern-Simons
path integral by going to its double branched cover, just as we have done in the orbifold theory on the boundary. 
The topology of the double branched cover can be understood as follows. Consider a minimal surface bounded
by each vortex and the boundary of the 3 ball, and cut open the 3-ball along the two minimal surfaces as shown
in Fig. \ref{3ball}. Gluing two copies of 3-balls across these minimal surfaces gives a solid torus of modulus $\tau$. Fig. \ref{fig:cyclea} and Fig. \ref{fig:cycleb} further illustrate this by circling the contractible and non-contractible cycles in the solid torus, respectively. 
 
The Chern-Simons path integral then gives $(\sqrt{\tau_2}/|\tau|)^c$, which is the $\gamma=1$ term in eq. 
 (\ref{EiInf}) . Other terms in eq. (\ref{EiInf})  are then generated by the monodromy transformations on $x$.
Therefore, the Eisenstein series (\ref{EiInf})  can be interpreted as a sum over rational 2-tangles pair-wisely
 connecting the points $0$ and $\infty$ and 
the points $x$ and $1$.

On the other hand, $SL(2, {\mathbb Z})/\Gamma(2)$ exchanges $0, 1, x$, and $\infty$ on the sphere
and generates vortex configurations connecting different pairs of the 4 points. Therefore, 
the three Eisenstein series in (\ref{TwstCorrAverage}) can all be interpreted in terms of sums over
rational 2-tangles. 

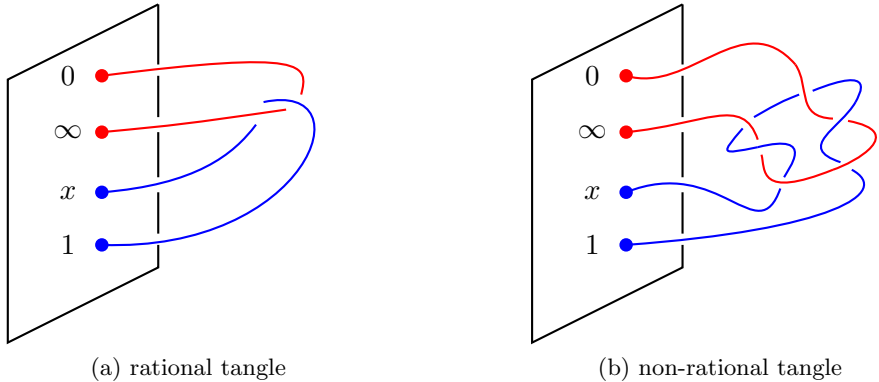
\begin{figure}
\centering
\captionsetup{width=1\linewidth}
\captionsetup[subfigure]{width=100pt}
\subfloat[rational tangle]{
\centering
\begin{tikzpicture}
\draw[thick, black] (2,3) -- (0,2) -- (0,-1.5) -- (2,-0.5) -- (2,-0.2);
\draw[thick, black] (2,-0.05) -- (2,0.5);
\draw[thick, black] (2,0.7) -- (2,1.3);
\draw[thick, black] (2,1.5) -- (2,2.05);
\draw[thick, black] (2,2.2) -- (2,3);
\draw[thick, red] (1.25,2.05) .. controls (4.0,2.4) and (4,2.2) .. (3.9,1.8);
\draw[thick, red]  (3.7,1.6) .. controls (2.3,1.4) .. (1.25,1.3);
\draw[thick, blue] (1.25,-0.2) .. controls (4,-0.3) and (4.8,2) .. (3.4,1.7);
\draw[thick, blue] (1.25,0.5) .. controls (2.5,0.6) and (3,1) .. (3.3,1.4);
\draw[thick, red, fill=red] (1.25,2.05) circle (0.75mm);
\draw[thick, red, fill=red] (1.25,1.3) circle (0.75mm);
\draw[thick, blue, fill=blue] (1.25,0.5) circle (0.75mm);
\draw[thick, blue, fill=blue] (1.25,-0.2) circle (0.75mm);
\node at (0.8,2.05) {0} node at (0.8,1.3) {$\infty$} node at (0.8,0.5) {$x$} node at (0.8,-0.2) {1};
\end{tikzpicture}
\label{rtan}
}\hspace{50pt}
\subfloat[non-rational tangle]{
\begin{tikzpicture}
\draw[thick, black] (2,3) -- (0,2) -- (0,-1.5) -- (2,-0.5) -- (2,-0.2);
\draw[thick, black] (2,-0.05) -- (2,0.5);
\draw[thick, black] (2,0.7) -- (2,1.3);
\draw[thick, black] (2,1.5) -- (2,2.05);
\draw[thick, black] (2,2.2) -- (2,3);
\draw[thick, red] (1.25,2.05) .. controls (2.1,1.8) and (2.8,3) .. (3.5,2.2);
\draw[thick, red]  (3.5,2.2) .. controls (3.7,1.9) and (3.5,1.5) .. (4,1.48);
\draw[thick, red]  (1.25,1.3) .. controls (2.3,1.4) and (2.8,1.8) .. (3,1.2);
\draw[thick, red]  (3.05,1) .. controls (3.1,0.8) and (3,0.4) .. (4.0,0.8);
\draw[thick, red]  (4.0,0.8) .. controls (4.3,0.9) and (5,1.3) .. (4.2,1.45);
\draw[thick, blue] (1.25,-0.2) .. controls (4,0.0) and (4.8,0.5) .. (4.25,0.8);
\draw[thick, blue] (4.1,0.9) .. controls (3.5,1.1) and (4.1,1.35) .. (4.3,1.7);
\draw[thick, blue] (4.3,1.7) .. controls (4.35,1.8) and (4.6,2.2) .. (3.73,1.85);
\draw[thick, blue] (3.55,1.8) .. controls (3.3,1.7) .. (2.95,1.5);
\draw[thick, blue] (1.25,0.5) .. controls (2.5,1) and (3,-0.3) .. (3.3,0.55);
\draw[thick, blue] (3.35,0.7) .. controls (3.4,0.8) and (3.8,1.3) .. (3,1.1);
\draw[thick, blue] (3,1.1) .. controls (2.5,1.0) .. (2.8,1.35);
\draw[thick, red, fill=red] (1.25,2.05) circle (0.75mm);
\draw[thick, red, fill=red] (1.25,1.3) circle (0.75mm);
\draw[thick, blue, fill=blue] (1.25,0.5) circle (0.75mm);
\draw[thick, blue, fill=blue] (1.25,-0.2) circle (0.75mm);
\node at (0.8,2.05) {0} node at (0.8,1.3) {$\infty$} node at (0.8,0.5) {$x$} node at (0.8,-0.2) {1};
\end{tikzpicture}
\label{nrtan}}
\caption{(a) an example of a rational tangle. (b) an example of an irrational tangle which cannot be generated by a finite sequence of exchanges of points on the boundary.}
\label{fig_r_nr_tangle}
\end{figure}

This raises the question on why configurations of vortices are restricted to be those of rational tangles. 
For example, a tangle configuration depicted in Fig. \ref{nrtan} cannot be generated by the monodromy group
on the trivial tangle, and there is no corresponding term in the Eisenstein series. This question can be answered as follows.

Since all rational 2-tangles are homeomorphic to the trivial 2-tangle, their double branched covers
are solid tori with the modulus $\tau$ related to each other by ${\Gamma_\infty} \backslash \Gamma(2)$. 
It turns out that the converse is also true.
There has been an extensive study of branched covers over knots and tangles in three dimensions. 
In particular, it was proven by Lemmas 4.4 and 4.5 in \cite{Hodgson} that,
if a double branched cover over a 2-tangle is a solid torus, the tangle must be rational.\footnote{We 
thank Toshitake Kohno, Kimihiko Motegi,
Makoto Sakuma, and Akira Yasuhara for explaining this to us.} Therefore, 
a 2-tangle is rational if and only if its double branched covering is a solid torus. Since
a three-dimensional manifold bounded by a genus-one surface admits
a hyperbolic metric if and only if it is a solid torus \cite{Maloney:2007ud}, 
we can also say that the Eisenstein series (\ref{EiInf}) is a sum over
vortex configurations such that their double branched covers allow hyperbolic metrics. 

\section{Averaging correlation functions of $\Z_N$ orbifolds}\label{sec:ZNCor}
Let us now sketch the computation for the averaged 4-point correlation functions of $\Z_N$ orbifolds. We will consider the correlator
\be\label{ZNcorr}
\langle \sigma_-(\infty)\sigma_+(1)\sigma_-(x)\sigma_+(0)\rangle\ .
\ee
There are now $N$ sectors which we denote by $k=0,1,\ldots N-1$. In the above we use the notation that $\sigma_+$ is in the $k^{\text{th}}$ sector, and $\sigma_-$ in the $(N-k)^{\text{th}}$ sector. To compute (\ref{ZNcorr}), we again go to a covering surface, which in this case turns out to be genus $N-1$. More precisely, this genus will have $\mathbb Z_N$ symmetry. In particular this means that we end up in a one-dimensional sublocus of the moduli space, parametrized by the cross ratio $x$. For the case of $\Z_3$, the relation between $x$ and the Riemann period matrix is for instance given by \cite{Calabrese:2009ez, Cho:2017fzo,Cardy:2017qhl}
\be
\Omega = \begin{pmatrix} 2&-1\\-1&2\end{pmatrix} 
\frac i{\sqrt{3}} \frac{{}_2F_1(\frac23,\frac13;1;1-x)}{{}_2F_1(\frac23,\frac13;1;x)}.
\ee
More generally, we want to define a `fake torus modulus' as \cite{Dixon:1986qv}
\be\label{tauN}
\tau_N(x) = \frac i{2\sin(\pi k/N)}\frac{{}_2F_1(\frac kN,1-\frac kN;1;1-x)}{{}_2F_1(\frac kN,1-\frac kN;1;x)}.
\ee
Even though for $N>2$ the covering surface is not a torus, the resulting expression for (\ref{ZNcorr}) looks very much like the partition on a torus with modulus $\tau_N$, and the symmetry group acts like a subgroup of $SL(2,\Z)$ on $\tau_N$. This is why we call $\tau_N$ a fake torus modulus. For $N=2$, $\tau_2$ is indeed simply the modulus of the covering torus.

In principle this genus $g=N-1$ surface has mapping class group $Sp(2g,\Z)$. However, since we are restricting to the symmetric sublocus, our symmetry group will be much smaller. We want to construct the action of the monodromy group of the sphere with four punctures on $\tau_N$. This group is freely generated by the monodromies around $0$ and $1$,
\be
A: x \mapsto e^{2\pi i}x \ , \qquad B:  x \mapsto 1- e^{2\pi i}(1-x)\ .
\ee
As discussed in section~\ref{s:corrfcts}, we want to think of this as the principal congruence subgroup $\Gamma(2)$, which is indeed freely generated by two generators, for instance (\ref{STS}) and (\ref{TT}).

We now want to consider their action on the fake modulus $\tau_N$. 
Using (\ref{tauN}) and the known monodromies of hypergeometric functions we find that they act on the fake torus modulus as
\be\label{tauNaction}
A: \tau_N \mapsto \tau_N - 1 \qquad B: \tau_N \mapsto \frac{ \tau_N}{4 \sin^2(\pi k/N) \tau_N +1}
\ee
First we note that if $4\sin^2(\pi k/N)$ is integral, then $A$ and $B$ in (\ref{tauNaction}) generate a subgroup $G$ of $SL(2,\Z)$. In the cases we are most interested in, for $k=1$ these subgroup are given by
\be\label{ZNsubgroups}
\begin{array}{c|c|c|c}
\Z_2 & \Z_3 & \Z_4 & \Z_6\\
\hline
4&3&2&1\\
\hline
\Gamma(2) & \Gamma_0(3) & \Gamma_0(2) & SL(2,\Z)	
\end{array}
\ee 
where the second row gives the value of $4\sin^2(\pi/N)$. Note that for $\Z_2$ when working with $\tau_2$ rather than $\tau$ we use the fact that $\Gamma_0(4)\simeq \Gamma(2)$. Next we note that the group action of $\Gamma(2)$ is not faithful for $N=3, 4, 6$. This means that even though the groups $G$ are still generated by $A$ and $B$, they are no longer freely generated, but rather satisfy some relations $R(A,B)=1$ and can therefore be written as a quotient $G=\Gamma(2)/R$. Physically this means that there are inequivalent tangle configurations which nonetheless give the same contribution to the correlation function. In the notation of (\ref{tauNaction}), the relations $R$ are: $(AB)^3=1$ for $k=1, N=3$; $(AB)^2=1$ for $k=1, N=4$; and $(AB)^3 = (BAB)^2 = 1$ for $k=1, N=6$.

The correlation function (\ref{ZNcorr}) is schematically given by \cite{Dixon:1986qv}
\be\label{ZNCorrZ}
Z \sim \frac{|x(1-x)|^{-2(k/N)(1-k/N)}}{V_\Lambda \sin(\pi k/N)|F(x)|^2}
	\sum_{p\in \Lambda^*, v\in \Lambda_c} e^{-2\pi i(f_{\epsilon_2}-f_{\epsilon_1})p} q^{(p+v/2)^2/2}\bar q^{(p-v/2)^2/2}
\ee
where $q= e^{2\pi i \tau_N}$. See \cite{Dixon:1986qv} for an explanation of the various quantities appearing this expression.
In view of the above, we expect $Z$ to be a modular form of $G$. To compute its average over the moduli space of $\Lambda$, we use the same tricks as in the previous section: establish that (\ref{ZNCorrZ}) satisfies the analogue of (\ref{mainDE4pt}), which after integrating over the K\"ahler moduli implies that it is an eigenfunction of the Laplace operator and can therefore be expressed as a linear combination of Eisenstein series with respect to an appropriate modular group. We expect the only change to be the number of cusps and therefore the dimension of the space of Eisenstein series.

We therefore conclude that the averaged correlation functions of the ${\mathbb Z}_N$ orbifold with $N=3, 4, 6$
are also expressed as sums over vortices in rational tangle configurations. As in the ${\mathbb Z}_2$ case,
a 2-tangle is rational if and only if its $N$-fold branched cover is a genus-$(N-1)$ handle body.
This statement can be shown by using the Smith conjecture, whose proof was outlined by 
W. Thurston as explained by J. Morgan in \cite{smith}. Thus, one may be tempted to claim that
the averaged correlation functions are equal to sums of 
vortex configurations such that their double branched covers allow hyperbolic metrics. 
Unfortunately, this is not quite right as there are non-handlebodies with 
hyperbolic metrics \cite{Yin:2007at}. This 
is also an issue with the higher genus partition functions in the $T^c$ case \cite{Afkhami-Jeddi:2020ezh, Maloney:2020nni}.

Using (\ref{ZNCorrZ}), we can also extract 3-point functions. These go beyond what we found in the $\Z_2$ case: there, the only non-vanishing 3-point function is between two twist fields and the vacuum, giving a 2-point function. Now however we can extract a 3-point function between $\sigma_+$, $\sigma_+$ and $\sigma_{--}$, where $\sigma_{--}$ is the $(N-2k)^{\text{th}}$ sector twist field. This 3-point function can be obtained from (\ref{ZNcorr}) by extracting the leading term in the $x\to\infty$ limit, which gives \cite{Dixon:1986qv}
\be
\langle \sigma_+ \sigma_+ \sigma_{--}\rangle 
=\sqrt{V_\Lambda |\tan \pi k /N|}\frac{\Gamma^2(\frac12+\frac12|1-2k/N|)}{\Gamma(|1-2k/N|)}\sum_{\vec v} e^{-\frac{\pi v^2}{|\sin 2\pi k/N|}}\ .
\ee
Note that even though we extracted the 3-point function from a pinching limit of the 4-point function, the fake torus does not become degenerate, since $\tau_N$ retains a finite imaginary part in the limit. In particular this explains why the 3-point function retains an infinite sum of lattice vectors.
We note that the 3-point function is non-vanishing. This agrees with the fact that in the bulk there is a configuration of one tangle each connecting $\sigma_+$ to $\sigma_{--}$.

\section{Puzzles about non-factorized amplitudes}
\label{sec:Nonfac}

We can also consider products of partition functions and correlation functions and average them over the Narain moduli space.
There are some puzzles. 

For the Narain theory studied by \cite{Afkhami-Jeddi:2020ezh, Maloney:2020nni}, 
the bulk dual of the average of the product of the partition functions is a sum of three-dimensional hyperbolic geometries bounded
by two genus-one surfaces. In particular, one can consider a connected geometry given by a solid torus with another solid torus dug out 
in the middle. This gives a non-factorizable contribution to the product of the partition functions. 

Suppose we apply the same rule to the orbifold theory to compute the average of the product of its partition functions. 
In the CFT side, the partition function receives contributions from the untwisted sector and the twisted sectors. 
Since the twisted sector partition function is independent of the Narain moduli,
the average of the product of the twisted sector partition function and the partition function of
 either the untwisted or twisted sector  
should factorize. Namely, for $Z_{T^c/{\mathbb Z}_N}$ defined by (\ref{Zorb}), 
\be\label{FactorizationPuzzle}
\langle Z_{T^c/{\mathbb Z}_N} Z_{T^c/{\mathbb Z}_N} \rangle - \langle Z_{T^c/{\mathbb Z}_N}\rangle 
\langle Z_{T^c/{\mathbb Z}_N}\rangle
= \frac{1}{N^2} \left( \langle Z_{(0,0)}Z_{(0,0)}\rangle - \langle Z_{(0,0)}\rangle \langle Z_{(0,0)}\rangle  \right) \ .
\ee

 However, on the bulk side, there are connected geometries bounded by two genus-one surfaces in twisted sectors.
For example, consider a solid torus with the ${\mathbb Z}_2$ gauge vortex threading through it. If we dig out another solid torus
inside of the solid torus, removing the gauge vortex with it, we obtain two genus-one surfaces in the same twisted sector. If we do not remove the gauge vortex, we obtain one boundary in the untwisted sector and another in the twisted sector. We have not been able to show that
these contributions combine with those from the untwisted sector to give rise to the right-hand side of 
(\ref{FactorizationPuzzle}) with the factor $1/N^2$.

There are related puzzles for correlation functions.
Consider, for example, the two-point function of twist operators on 
the sphere. Whether it is normalized (divided by the sphere partition function)
or not, the two-point funciton is independent of the Narain moduli.
Therefore, the average of the product of the two-point functions should factorize.
However, in the bulk, there are connected geometries bounded by two
spheres; if we consider a solid ball and dig out another solid ball in its middle,
the resulting geometry has a hyperbolic structure and is bounded by two spheres. 
We can then add two vortices
with two ends on each sphere. Each vortex can either end on the same sphere or 
go between the spheres. Unless the sum over such vortex configurations in the
connected geometry cancel out, the bulk amplitude does not factorize into the product
of two-point functions. There are similar and more elaborate puzzles for higher point functions.
We hope to revisit these puzzles in future.

\section{Averaging over K3 and Calabi-Yau moduli spaces?}
\label{sec:K3Gen}

In this section, we will attempt to generalize the previous discussion to averaging over K3 and Calabi-Yau threefold (CY3) moduli spaces. Because these CFTs are interacting, we no longer will be able to use the differential equation (\ref{eq:LaplacianWeWant}) to determine the average partition function. However, a na\"ive guess may be that the result of the computation is the same -- the average partition function may be the Poincar\'e sum of the $\mathcal{N}=2$ or $\mathcal{N}=4$ vacuum character, or the moduli-independent pieces of the partition function. In this section we will show that the former guess is not correct, as it leads to a spectrum that is not positive definite. We also show the latter guess is not correct for the case of Calabi-Yau threefolds, and set up a similar calculation for K3.

In addition to potential relations to gravity-like theories, calculating an averaged K3 or Calabi-Yau partition function would shed light on questions about how a ``typical" K3 or Calabi-Yau CFT behaves. For instance, the distribution of rational points in K3 and Calabi-Yau CFTs is an open and interesting problem \cite{Gukov:2002nw}. If one could calculate the twist gap of the average K3 partition function under the $\mathcal{N}=4$ superconformal algebra, that may be an indication of how dense rational points are (although we pause to note that the vanishing of the twist gap is neither a necessary nor sufficient condition for rational points being dense \cite{Benjamin:2020flm}). 

\subsection{Review of $\mathcal{N}=4$ representation theory}

Before we perform the Poincar\'e sums, let us first review the representation theory of the small $\mathcal{N}=4$ superconformal algebra at $c=6$. The algebra has two massless (BPS) representations and one family of massive (non-BPS) representations. Following the notation of \cite{Eguchi:2003ik}, we will refer to the characters of the two BPS representations as $\chi_G$ and $\chi_M$, where $\chi_G$ is the character containing the unique NS ground state. Finally, we refer to the massive characters $\chi^p$ as the non-BPS representation with weight $h=\frac{p^2}{2}+\frac18$ above the unitarity bound, i.e.
\be
\chi^p(\tau, z) = \(\chi^G(\tau, z) + 2\chi^M(\tau, z)\)q^{\frac{p^2}2+\frac18}.
\ee
Explicit expressions for the characters were computed by Eguchi and Taormina in \cite{Eguchi:1987sm, Eguchi:1987wf} in both the NS and Ramond sectors. However, for our purposes the modular kernels are enough, which were computed in \cite{Eguchi:1987wf, Eguchi:1988af}. We reproduce them in Appendix \ref{sec:c6kernels} in Eqns (\ref{eq:Skernels}) and (\ref{eq:Tkernels}). We have chosen to work in the $\tilde R$ spin structure (defined as the trace in the Ramond sector with a $(-1)^F$ inserted) for convenience. 

\subsection{Poincar\'e sum of the K3 vacuum character}

We now would like to consider the Poincar\'e sum of the vacuum character as a possible candidate for the average partition function of K3 CFTs:\footnote{The phase on the RHS of (\ref{eq:FirstAttemptK3}) is to take into account the chemical potential in the R-symmetry of the $\mathcal{N}=4$ superconformal algebra. See \cite{Datta:2021ftn} for a recent paper grading by the chemical potentials of the $U(1)^c$ algebra in the Narain ensemble.}
\be
Z^{\text{average K3}}(\tau, \bar\tau, z, \bar z) \stackrel{?}{=} \sum_{\gamma \in \Gamma_\infty \backslash SL(2,\mathbb Z)} e^{-2\pi i \(\frac{cz^2}{c\tau+d} - \frac{c \bar z^2}{c\bar\tau+d}\)} |\chi_G(\gamma \tau, \gamma z)|^2.
\label{eq:FirstAttemptK3}
\ee
In this section, we will show that the sum (\ref{eq:FirstAttemptK3}) cannot reproduce the average K3 partition function for two reasons. First, its Witten index does not match the K3 Witten index of 24. Second, the non-BPS spectrum of (\ref{eq:FirstAttemptK3}) is not positive definite. 

To calculate the Witten index of (\ref{eq:FirstAttemptK3}), we compute it in the $\tilde R$ sector and set $z=\bar z =0$. In the $\tilde R$ sector, $\chi_G(\gamma \tau, 0) = -2$ for all $\gamma$. This then gives
\be
Z^{\text{average K3}}(\tau, \bar\tau, 0, 0) \stackrel{?}{=} \sum_{\gamma \in \Gamma_\infty \backslash SL(2,\mathbb Z)} 4.
\label{eq:FirstAttemptK3Witten}
\ee
The sum in (\ref{eq:FirstAttemptK3Witten}) diverges of course. However it can be regulated by taking the real analytic Eisenstein series 
\be
E(s, \tau, \bar\tau) = \sum_{\gamma \in \Gamma_\infty \backslash SL(2,\mathbb Z)} \frac{y^s}{|c\tau+d|^{2s}}
\ee
and continue $s\rightarrow 0$. From the explicit form of the Fourier expansion of the Eisenstein series, we get
\be
E(0, \tau, \bar\tau) = 1.
\ee
Thus the Witten index in (\ref{eq:FirstAttemptK3Witten}), with the regularization scheme as described above, is $4$, which does not match the desired answer of $24$.

The second problem with (\ref{eq:FirstAttemptK3}) is more subtle. We will show that the spectrum of non-BPS states it has is not positive-definite. This is analogous to the negativity found in \cite{Benjamin:2019stq} for the Poincar\'e sum of the Virasoro vacuum character
\be
Z^{\text{MWK}}(\tau, \bar\tau) = \sum_{\gamma \in \Gamma_\infty \backslash SL(2,\mathbb Z)} |\chi_\text{vacuum}^{\text{Virasoro}}(\tau)|^2,
\label{eq:MWKker}
\ee
which was considered in \cite{Maloney:2007ud, Keller:2014xba}. The non-BPS spectrum of (\ref{eq:FirstAttemptK3}) can be read off from the modular kernels
\be
\rho^{\text{average~K3}}(p, \bar p) \stackrel{?}{=} \sum_{\gamma \in \Gamma_\infty \backslash SL(2,\mathbb Z)} K^\gamma(p) K^\gamma(\bar p)
\label{eq:rhok3}
\ee
where the kernel $K^\gamma(p)$ is defined as
\be
\chi^G(\gamma \tau, \gamma z) =e^{2\pi i \frac{cz^2}{c\tau+d}} \left[ \text{BPS states} + \int_0^\infty dp K^\gamma(p) \chi^p(\tau, z)\right].
\ee
In (\ref{eq:rhok3}) we have by convention labelled the weights of the states by their Liouville momenta $p$ instead of their conformal weights $h$, which are related by $h=\frac{p^2}{2}+\frac18$.

Following \cite{Benjamin:2019stq}, we will consider the limit of (\ref{eq:rhok3}) in the limit of $p\rightarrow 0$, and $\bar p \rightarrow\infty$, i.e. the limit of low twist and high spin. It is important that we first take the limit of low twist and second take the limit of high spin. In the limit as $p\rightarrow 0$, the kernel scales differently with $p$ depending on the elements $\begin{pmatrix} a & b \\ c & d \end{pmatrix} \in SL(2,\mathbb Z)$. In Appendix \ref{sec:c6kernels}, we show that in this limit, 
\begin{align}
K^S(p) &= 2\sinh(\pi p)\tanh(\pi p) = \mathcal{O}(p^2) \nn\\
K^{ST^2S}(p) &= \sqrt{2} e^{-i \pi \(\frac{p^2}2 + \frac18\)}\(1 - \frac{7\pi^2}{8}p^2 +  \frac{113\pi^4}{384} p^4 + \mathcal{O}(p^6)\),
\label{eq:KSKSTSscale}
\end{align}
where every term in the series expansions in (\ref{eq:KSKSTSscale}) are real numbers.
Thus we see that in this limit, the modular transform $ST^2S$ dominates over the $S$ transformation\footnote{In fact it can be shown that the scaling (\ref{eq:KSKSTSscale}) with $p$ is the same for any $\begin{pmatrix} a & b \\ c & d \end{pmatrix} \in SL(2,\mathbb Z)$ element with $c>1$, and $ST^2S$ is the most dominant amongst such transformations at large spin.}. Moreover due to the phase, the spectrum obtained from the modular transformation $ST^2S$ has sign $(-1)^j$, and is therefore not positive definite. In this limit of small twist and large spin, then, the non-BPS spectrum is not positive definite for odd spins. Thus, like in the case of the Virasoro algebra, but \emph{unlike} the case of the $U(1)^c$ algebra, the spectrum obtained from Poincar\'e sum of the vacuum character is not positive-definite. Since each individual K3 spectrum is positive-definite, its average must also be positive-definite, and (\ref{eq:FirstAttemptK3}) cannot be interpreted as an average. 

We pause here to clarify a potentially confusing point. Since we chose to compute the partition function in the $\tilde R$ spin structure, namely 
\be
Z^{K3}(\tau, \bar\tau, z, \bar z) = \Tr_{\text{RR}} \( (-1)^F q^{L_0-\frac{c}{24}} y^{J_0} \bar{q}^{\overline{L_0}-\frac c{24}} \bar{y}^{\overline{J_0}}\),
\label{eq:zdefrm}
\ee
and (\ref{eq:zdefrm}) is not a positive definite due to the $(-1)^F$, it may be unclear what we mean by the spectrum not being positive definite. However, (\ref{eq:zdefrm}) must be positive-definite in the following sense. If we expand (\ref{eq:zdefrm}) into the $c=6$ small $\mathcal{N}=4$ characters in the $\tilde R$ spin structure, the overlap coefficients must be positive. In other words, we write
\be
Z^{K3}(\tau, \bar\tau, z, \bar z) = \int dh d{h} dQ d\bar{Q} \rho(h, \bar h, Q, \bar Q) \chi^{h, Q}(\tau, z) \chi^{\bar h, \bar Q}(\bar\tau, \bar z),
\label{eq:charexp}
\ee
where
\be
\chi^{h, Q}(\tau, z) = \Tr_{(h,Q)~\text{representation}, R} \( (-1)^F q^{L_0-\frac{c}{24}} y^{J_0} \)
\ee
are the $\tilde R$ characters, which are related to the other characters by spectral flow. In (\ref{eq:charexp}) we have combined all BPS and non-BPS characters abstractly into a single integral. The quantity $\rho(h, \bar h, Q, \bar Q)$ must be non-negative for all $h, \bar h, Q, \bar Q$, and indeed this quantity is invariant under spectral flow. When we compute the Poincar\'e series of the $\mathcal{N}=4$ vacuum character, we extract a non-positive-definite density $\rho$, so it cannot be the averaged K3 partition function. 

\subsection{Poincar\'e sum of K3 BPS states}
\label{sec:k3bpssec}

In Sections \ref{sec:tcz2} and \ref{sec:tczn} we showed that to reproduce the average of orbifolds of free bosons, one must do a Poincar\'e sum of not only the vacuum character, but all moduli-independent characters. A natural generalization to the case of K3, then, is to not only do the Poincar\'e sum of the $\mathcal{N}=4$ vacuum character, but all BPS states. Fortunately, in the case of K3 (unlike for higher dimensional Calabi-Yau manifolds), there exists a generic BPS spectrum shared by almost all points in moduli space. This was first explored in \cite{Ooguri:1989fd}, and is because of the following:

Let us consider the most general possible spectrum of a K3 sigma model in terms of the $\mathcal{N}=4$ characters.
\begin{align}
Z^{K3}(\tau, \bar\tau, z, \bar z) &= \chi_G(\tau, z)\chi_G(\bar\tau, \bar z) + 20\chi_M(\tau, z)\chi_M(\bar\tau, \bar z) \nn\\
&+ \sum_{h=1}^{\infty} N_h q^h \(\chi_G(\tau, z) + 2\chi_M(\tau, z)\)\chi_M(\bar\tau, \bar z) + c.c. \nn\\
&+ \sum_{h=1}^{\infty} M_h q^h \(\chi_G(\tau, z) + 2\chi_M(\tau, z)\)\chi_G(\bar\tau, \bar z) + c.c. \nn\\
&+ \sum_{h, \bar h>0} N_{h, \bar h} q^h \bar{q}^{\bar h} \(\chi_G(\tau, z) + 2\chi_M(\tau, z)\)\(\chi_G(\bar\tau, \bar z) + 2\chi_M(\bar \tau, \bar z)\).
\label{eq:K3gen}
\end{align}
The first line of (\ref{eq:K3gen}) is the half-BPS states; the second and third lines are the quarter-BPS states; and the final line is the non-BPS states. By unitarity, $N_h, M_h$, and $N_{h, \bar h}$ are all non-negative integers.

The elliptic genus of K3 is obtained by evaluating (\ref{eq:K3gen}) at $\bar z =0$ in the $\tilde R$ spin structure, and constraints $N_h - 2M_h$ for all $h=1, 2, \ldots$. The assumption in \cite{Ooguri:1989fd} was that all $M_h=0$, because any nonzero $M_h$ would correspond to an enlargement of the chiral algebra of the sigma model, and it is expected that a generic point in moduli space should only have the small $\mathcal{N}=4$ superconformal algebra. Therefore, this uniquely determines the $N_h$'s as
\be
Z^{EG}(q,y) = -2 \chi^G(q, y) + 20 \chi^M(q,y) + \sum_{h=1}^{\infty} N_h(\chi^G(q,y) + 2\chi^M(q,y))q^h.
\label{eq:EGK3}
\ee
The first few values of $N_h$ obtained from (\ref{eq:EGK3}) are
\be
\sum_{h=1}^{\infty} N_h q^h = 90q + 462q^2 + 1540q^3 + \ldots.
\ee
We pause to note that these degeneracies appear to have an interesting connection with the sporadic Mathieu group $M_{24}$ \cite{Eguchi:2010ej}. 

We now then would like to explore the following proposal:
\begin{align}
&Z^{\text{average K3}} \stackrel{?}{=}  \sum_{\gamma \in \Gamma_\infty \backslash SL(2,\mathbb Z)} e^{-\frac{2\pi i c z^2}{c\tau+d}+\frac{2\pi i c {\bar z}^2}{c\bar\tau+d}}
\nn\\&~~~\times\(|\chi_G(q_\gamma,y_\gamma) + 2\chi_M(q_\gamma,y_\gamma)|^2 - 24|\chi_M(q_\gamma,y_\gamma)|^2 + Z_{\text{EG}}(q_\gamma, y_\gamma)\chi_M(\bar q_\gamma, \bar y_\gamma) + \chi_M(q_\gamma, y_\gamma)Z_{\text{EG}}(\bar q_\gamma, \bar y_\gamma) \). 
\label{eq:K3fullshebang}
\end{align}
The sum on the RHS of (\ref{eq:K3fullshebang}) represents the K3 BPS spectrum at generic points in the moduli space. Although we are unable to get an exact expression for the sum in (\ref{eq:K3fullshebang}), we believe its spectrum retains the negativity from the Poincar\'e sum of the vacuum character. In the next sections we will do the same type of sum, but for Calabi-Yau threefolds and show that this is indeed the case there. 

In this and the previous subsections, we were unable to calculate the average K3 partition function as a Poincar\'e sum. However, we can of course compute the average when we only integrate over a special sublocus in moduli space, instead of the full 80 dimensions. For instance, we can integrate over the 16-dimensional sublocus of moduli space where the K3 surface is realized as a $T^4/\mathbb Z_2$ orbifold, called the Kummer locus. The average over this locus in moduli space will simply be the $\mathbb{Z}_2$ orbifold at $c=4$ computed in Section \ref{sec:tcz2}, combined with the partition function of four fermions. 

\subsection{Review of extended $\mathcal{N}=2$ representation theory}

In this section we briefly review the extended $\mathcal{N}=2$ superconformal algebra \cite{Odake:1988bh, Odake:1989ev}. The algebra is defined for central charge $c=3\hat c, ~\hat c\in\mathbb N$, and is the algebra obtained on the string worldsheet with target space Calabi-Yau $\hat c$-fold. 

In general there will be $\hat c$ short multiplets and $\hat c-1$ families of long multiplets. We will follow the conventions in \cite{Eguchi:2003ik} and label the multiplets by their highest-weight $U(1)$ charge in the NS sector. We will label the short multiplets as $\chi_G(\tau, z)$ for the unique vacuum multiplet, and $\chi_M^Q(\tau,z)$, for the non-vacuum short multiplets with charge $Q$ in the NS sector. $Q$ runs from $Q = \pm 1, \ldots, \pm \frac{\hat c - 1}2$ for odd $\hat c$, and $Q = \pm 1, \ldots, \pm (\frac{\hat c}2-1), \frac{\hat c}2$ for even $\hat c$. There are $\hat c-1$ families of long multiplets. We will label them as $\chi_Q^p$, where $Q$ is their charge in the NS sector and $p$ is their Liouville momentum. For the long multiplets, $Q = -\frac{\hat c-3}2, \ldots, \frac{\hat c-1}2$ for odd $\hat c$, and $Q = -\frac{\hat c}2+1, \ldots, \frac{\hat c}2-1$ for even $\hat c$. Finally the Liouville momentum $p$ is related to the weight by $h = \frac{p^2}2 + \frac{Q^2}{2(\hat c-1)} + \frac{\hat c-1}8$ in the NS sector. 

\subsection{Poincar\'e sum of $\mathcal{N}=2$ vacuum character}

The first question we have is what is the Poincar\'e sum of the vacuum character, and can the average of some Calabi-Yau moduli space be interpreted as this Poincar\'e sum? 
\be
Z^{\text{average CY $\hat c$-fold}}(\tau, \bar\tau, z, \bar z) \stackrel{?}{=} \sum_{\gamma \in \Gamma_\infty \backslash SL(2,\mathbb Z)} e^{-2\pi i \(\frac{cz^2}{c\tau+d} - \frac{c \bar z^2}{c\bar\tau+d}\)} |\chi_G^{\hat c}(\gamma \tau, \gamma z)|^2.
\label{eq:FirstAttemptCY}
\ee 
In Appendix \ref{app:mwkcy}, we will show that this sum (\ref{eq:FirstAttemptCY}) is not positive definite for $c>3$, which means it cannot be interpreted as the average of unitary CFTs. 

We have thus shown that the Poincar\'e sum of the vacuum character of the extended $\mathcal{N}=2$ superconformal algebra does not have a positive definite spectrum for $c>3$. It was previously shown in \cite{Benjamin:2019stq} that the Poincar\'e sum of the Virasoro and $\mathcal{N}=1$ super-Virasoro conformal algebras are also not positive definite for $c>1$ and $c>\frac32$ respectively, and it was shown in \cite{Alday:2020qkm} that the Poincar\'e sum of the vacuum character of the $\mathcal{W}_N$ algebra for $c>N-1$ is not positive definite. Furthermore these negativities persist when we do an $SL(2,\mathbb Z)$ sum with a different regularization, such as the Rademacher regulator \cite{Alday:2019vdr}. This then leads us to a conjecture:

\noindent\fbox{%
    \parbox{\textwidth}{%
{\bf Conjecture:} A regularized $SL(2,\mathbb Z)$ sum of the vacuum character of any chiral algebra with $c>c_{\text{crit}}$ does not have a positive-definite spectrum.
    }%
}

\subsection{Poincar\'e sum of CY3 BPS states}

Similar to the K3 case, we can try to refine the above calculation by performing a Poincar\'e sum not just of the vacuum character, but all moduli-independent pieces in the conformal manifold. As in Sec \ref{sec:k3bpssec}, a natural guess then would be to sum over all BPS states in the theory. Unfortunately, at sufficiently high dimension, the half-BPS spectrum of a Calabi-Yau manifold does not determine its generic quarter-BPS spectrum. If we focus on Calabi-Yau threefolds, the representation theory is constraining enough that it does. In fact, assuming that the chiral algebra does not enhance, a generic point in a CY3 moduli space has no quarter-BPS highest weight states \cite{Keller:2012mr}. The generic BPS spectrum given a fixed $h^{1,1}$ and $h^{2,1}$ is given by
\be
Z^{\text{BPS}} = \chi^0 \overline{\chi^0} + h^{1,1}\(\chi^1 \overline{\chi^1} + \chi^{-1} \overline{\chi^{-1}}\)+ h^{2,1}\(\chi^1 \overline{\chi^{-1}} + \chi^{-1} \overline{\chi^{1}}\).
\label{eq:ZBPSCy3Top}
\ee
We can then ask the question: Is the average Calabi-Yau threefold partition function with Hodge numbers $h^{1,1}, h^{2,1}$ given by the Poincar\'e sum of (\ref{eq:ZBPSCy3Top})? 

In Appendix \ref{app:mwkcy} we will show that the answer is again \emph{no}: such a sum again does not have a positive-definite spectrum. Therefore, the statement that the Poincar\'e sum of the moduli-independent pieces of an orbifold CFT giving the average does not generalize to Calabi-Yau averages. This negative result also makes us believe that it is unlikely (\ref{eq:K3fullshebang}) has a positive-definite spectrum.

Given the negative results in this section, it is natural to ask if there are other situations in which the Poincar\'e sum of a vacuum character can or cannot be interpreted as a positive-definite average of CFTs. In Appendix \ref{app:minmod}, we calculate the Poincar\'e sum of the Virasoro minimal model vacuum characters and show that in general, they cannot be written as positive linear combinations of minimal model CFTs. 

\section*{Acknowledgements}
We thank A. Adams, K. Bringmann, S. Collier, S. Kachru, T. Kohno, A. Maloney, J. Manschot, G. Moore, K. Motegi, K. Ono, B. Rayhaun, L. Rolen, M. Sakuma, and A. Yasuhara for very helpful discussions.  
We thank S. Collier, T. Hartman, and A. Maloney for very helpful comments on a draft.
The work of N.B. is supported in part by the Simons Foundation Grant No. 488653. 
The work of C.A.K. is supported in part by the Simons Foundation Grant No.~629215.
The work of H.O. is supported in part by
U.S.\ Department of Energy grant DE-SC0011632, by
the World Premier International Research Center Initiative,
MEXT, Japan, by JSPS Grant-in-Aid for Scientific Research 17K05407  and 20K03965,
and by JSPS Grant-in-Aid for Scientific Research on Innovative Areas
15H05895.
H.O. thanks the Aspen Center for Theoretical Physics, which is supported by
the National Science Foundation grant PHY-1607611,  where part of this work was done.

\appendix

\section{Poincar\'e sum of the $T^c/\mathbb Z_2$ vacuum character}\label{app:PoinVac}

In this appendix, we will compute the Poincar\'e sum of the vacuum character of the $T^c/\mathbb Z_2$ CFT, and show it does not reproduce the averaged $\mathbb Z_2$ orbifold partition function. Recall the vacuum character is given in (\ref{eq:CharOrb}) which we reproduce below:
\be
\chi^{\text{vac}}(\tau) = \frac12\left[\frac1{\eta(\tau)^c} + \frac{\eta(\tau)^c}{\eta(2\tau)^c}\right] = \frac12\left[\frac1{\eta(\tau)^c} + \frac{2^{\frac c2} \eta(\tau)^{\frac c2}}{\theta_2(\tau)^{\frac c2}} \right]. 
\ee
We can then write the modular sum of the vacuum character as a sum of four terms:
\begin{align}
\sum_{\gamma \in \Gamma_\infty\backslash SL(2,\mathbb Z)} |\chi^{\text{vac}}(\gamma\tau)|^2 &= \frac14 \sum_{\gamma \in \Gamma_\infty\backslash SL(2,\mathbb Z)} \left[\frac{1}{|\eta(\gamma\tau)|^{2c}} + \frac{|\eta(\gamma\tau)|^{2c}}{|\eta(2\gamma\tau)|^{2c}} + \frac{\eta(\gamma\tau)^c}{\eta(2\gamma\tau)^c \eta(-\gamma\bar\tau)^c} + \frac{\eta(-\gamma\bar\tau)^c}{\eta(-2\gamma\bar\tau)^c \eta(\gamma\tau)^c}\right].
\label{eq:MWKFOrb}
\end{align} 
The first of the four terms in (\ref{eq:MWKFOrb}) is essentially the calculation done in \cite{Afkhami-Jeddi:2020ezh, Maloney:2020nni}, and gives
\be
\frac14 \sum_{\gamma \in \Gamma_\infty\backslash SL(2,\mathbb Z)} \frac{1}{|\eta(\gamma\tau)|^{2c}}  = \frac14  \frac{E\(\frac c2, \tau, \bar\tau\)}{\tau_2^\frac{c}2 |\eta(\tau)|^{2c}}.
\label{eq:AlexEd}
\ee
The second of the four terms is more complicated. Let us first split the sum over $\Gamma_\infty\backslash SL(2,\mathbb{Z})$ into three terms:  a sum over $\Gamma_0(2)$, and the two cosets $\Gamma_0(2).S$ and $\Gamma_0(2).ST$ where $\Gamma_0(2)$ is subgroup of $SL(2,\mathbb Z)$ generated by $T$ and $ST^2S$. Finally we mod out all cosets on the left by the group generated by $T$. The reason we do this splitting is because $\frac{|\eta(\tau)|^{2c}}{|\eta(2\tau)|^{2c}}$ is modular invariant (with weight 0) under $\Gamma_0(2)$, but not $SL(2, \mathbb{Z})$. Thus
\begin{align}
\frac14 \sum_{\gamma \in \Gamma_\infty \backslash SL(2,\mathbb Z)} \frac{|\eta(\gamma\tau)|^{2c}}{|\eta(2\gamma\tau)|^{2c}}  &= \frac 14 \left[  \sum_{\gamma \in \Gamma_\infty\backslash \Gamma_0(2)} \frac{|\eta(\gamma\tau)|^{2c}}{|\eta(2\gamma\tau)|^{2c}}  +   \sum_{\gamma \in \Gamma_\infty \backslash \Gamma_0(2).S} \frac{|\eta(\gamma\tau)|^{2c}}{|\eta(2\gamma\tau)|^{2c}}   +   \sum_{\gamma \in \Gamma_\infty\backslash \Gamma_0(2).ST} \frac{|\eta(\gamma\tau)|^{2c}}{|\eta(2\gamma\tau)|^{2c}}  \right].
\label{eq:SecondFour}
\end{align}
Using the following identities
\begin{align}
\frac{|\eta(\tau)|^{2c}}{|\eta(2\tau)|^{2c}} &= 2^c \frac{|\eta(\tau)|^c}{|\theta_2(\tau)|^c}\nn\\
\frac{|\eta(-\frac1\tau)|^{2c}}{|\eta(-\frac2\tau)|^{2c}} &= 2^c \frac{|\eta(\tau)|^{2c}}{|\eta(\frac\tau2)|^{2c}} =2^c \frac{|\eta(\tau)|^c}{|\theta_4(\tau)|^c} \nn\\
\frac{|\eta(-\frac1{\tau+1})|^{2c}}{|\eta(-\frac2{\tau+1})|^{2c}} &= 2^c \frac{|\eta(\tau)|^{2c}}{|\eta(\frac{\tau+1}2)|^{2c}} = 2^c \frac{|\eta(\tau)|^c}{|\theta_3(\tau)|^c},
\label{eq:JacIdent}
\end{align}
we get
\begin{align}
\frac14 \sum_{\gamma \in \Gamma_\infty\backslash SL(2,\mathbb Z)} \frac{|\eta(\gamma\tau)|^{2c}}{|\eta(2\gamma\tau)|^{2c}}  = \frac 14 \left[ 2^c \frac{|\eta(\tau)|^c}{|\theta_2(\tau)|^c} \sum_{\gamma \in \Gamma_\infty \backslash \Gamma_0(2)} 1 +  2^c \frac{|\eta(\tau)|^c}{|\theta_4(\tau)|^c}  \sum_{\gamma \in \Gamma_\infty \backslash \Gamma_0(2).S} 1   +  2^c \frac{|\eta(\tau)|^c}{|\theta_3(\tau)|^c}  \sum_{\gamma \in \Gamma_\infty \backslash \Gamma_0(2).ST} 1 \right].
\label{eq:SecondFourv2}
\end{align}
The sums like $\sum_{\gamma \in \Gamma_\infty \backslash \Gamma_0(2)} 1$ are clearly divergent, but can be defined via analytic continuation of the non-holomorphic Eisenstein series. In particular, if we define
\be
E^{\Gamma_0(2)}(s, \tau, \bar\tau) := \sum_{\gamma\in\Gamma_\infty\backslash \Gamma_0(2)} {\text{Im}(\gamma\tau)}^s
\ee
we can take the limit $s\rightarrow0$ which turns out to be finite. Then (\ref{eq:SecondFourv2}) would reduce to
\begin{align}
\frac14\sum_{\gamma \in \Gamma_\infty \backslash SL(2,\mathbb Z)} \frac{|\eta(\gamma\tau)|^{2c}}{|\eta(2\gamma\tau)|^{2c}}  &= \frac {2^c}4 \Bigg[\frac{|\eta(\tau)|^c}{|\theta_2(\tau)|^c} E^{\Gamma_0(2)}(0, \tau, \bar\tau) + \frac{|\eta(\tau)|^c}{|\theta_4(\tau)|^c}  E^{\Gamma_0(2)}(0, -\frac1\tau, -\frac1{\bar\tau})   \nn\\& ~~~~~~~~~~+ \frac{|\eta(\tau)|^c}{|\theta_3(\tau)|^c}  E^{\Gamma_0(2)}(0, -\frac1{\tau+1}, -\frac{1}{\bar\tau+1}) \Bigg].
\label{eq:SecondFourv3}
\end{align}
In Equations (C.20) and (C.22) of \cite{Benjamin:2020zbs}, explicit expressions for all three terms in (\ref{eq:SecondFourv3}) are given. They are:
\begin{align}
E^{\Gamma_0(2)}(0,\tau,\bar\tau)&= 1 - \frac{\pi y}{6\log(2)} + \sum_{j=1}^{\infty} \frac{2\cos(2\pi j x) e^{-2\pi j y}(\sigma_1(2j) - 2\sigma_1(j))}{j \log(2)}\nn\\
E^{\Gamma_0(2)}\(0,-\frac1\tau,-\frac1{\bar\tau}\)&= \frac{\pi y}{12\log(2)} - \sum_{j=1}^{\infty} \frac{\cos(2\pi j x) e^{-2\pi j y}(\sigma_1(2j) - 2\sigma_1(j))}{j \log(2)} \nn\\
&~~~~~~~~~~~~~~~- \sum_{j=\frac12, \frac32, \cdots} \frac{\cos(2\pi j x) e^{-2\pi j y} \sigma_1(2j)}{j\log(2)}\nn\\
E^{\Gamma_0(2)}\(0,-\frac1{\tau+1},-\frac{1}{\bar\tau+1}\)&= \frac{\pi y}{12\log(2)} - \sum_{j=1}^{\infty} \frac{\cos(2\pi j x) e^{-2\pi j y}(\sigma_1(2j) - 2\sigma_1(j))}{j \log(2)} \nn\\
&~~~~~~~~~~~~~~~+ \sum_{j=\frac12, \frac32, \cdots} \frac{\cos(2\pi j x) e^{-2\pi j y} \sigma_1(2j)}{j\log(2)},
\end{align}
where $\tau = x+i y, \bar\tau = x-i y$.

Finally we need the last two terms in (\ref{eq:MWKFOrb}). Since they are complex conjugates of each other let us focus on the third term:
\be
\sum_{\gamma \in \Gamma_\infty \backslash SL(2,\mathbb Z)}  \frac{\eta(\gamma\tau)^c}{\eta(2\gamma\tau)^c \eta(-\gamma\bar\tau)^c}.
\label{eq:ThirdFourth}
\ee
Unlike the first two terms -- which transformed roughly as weight $(1/2, 1/2)$ and weight $(0,0)$ modular forms respectively, the expression in (\ref{eq:ThirdFourth}) transforms roughly as a weight $(0,1/2)$ modular form. Moreover, because the holomorphic and anti-holomorphic terms are no longer the same, we now have to worry about the phase that occurs when we do an $SL(2,\mathbb{Z})$ transformation (unlike in the first two cases). We still will split (\ref{eq:ThirdFourth}) into three pieces based on the elements' relation to $\Gamma_0(2)$ because the holomorphic part of (\ref{eq:ThirdFourth}) is invariant not under $SL(2,\mathbb{Z})$ but $\Gamma_0(2)$:
\begin{align}
\sum_{\gamma \in \Gamma_\infty \backslash SL(2,\mathbb Z)}  \frac{\eta(\gamma\tau)^c}{\eta(2\gamma\tau)^c \eta(-\gamma\bar\tau)^c} &= \Bigg[  \sum_{\gamma \in \Gamma_\infty \backslash \Gamma_0(2)} \frac{\eta(\gamma\tau)^{c}}{\eta(2\gamma\tau)^{c} \eta(-\gamma\bar\tau)^c}  +   \sum_{\gamma \in \Gamma_\infty \backslash \Gamma_0(2).S} \frac{\eta(\gamma\tau)^{c}}{\eta(2\gamma\tau)^{c} \eta(-\gamma\bar\tau)^c}   \nn\\
&~~~~~~~~~~~~~~~~~~~~~~~~~~~~~~+   \sum_{\gamma \in \Gamma_\infty \backslash \Gamma_0(2).ST} \frac{\eta(\gamma\tau)^{c}}{\eta(2\gamma\tau)^{c} \eta(-\gamma\bar\tau)^c}  \Bigg] \nn\\
&= \frac{2^{\frac c2} \eta(\tau)^{\frac c2}}{\theta_2(\tau)^{\frac c2}\eta(-\bar\tau)^c} \sum_{\gamma \in \Gamma_\infty \backslash \Gamma_0(2)} (s\bar\tau+d)^{-c/2} e^{2\pi i c \epsilon_1(s,d)} \nn\\
&~~~ + \frac{2^{\frac c2} \eta(\tau)^{\frac c2}}{\theta_4(\tau)^{\frac c2}\eta(-\bar\tau)^c} \sum_{\gamma \in \Gamma_\infty \backslash \Gamma_0(2).S} (s\bar\tau+d)^{-c/2} e^{2\pi i c \epsilon_2(s,d)}  \nn\\
&~~~ + \frac{2^{\frac c2} \eta(\tau)^{\frac c2}}{\theta_3(\tau)^{\frac c2}\eta(-\bar\tau)^c} \sum_{\gamma \in \Gamma_\infty\backslash \Gamma_0(2).ST} (s\bar\tau+d)^{-c/2} e^{2\pi i c \epsilon_3(s,d)}
\label{eq:ThirdFourthv2}
\end{align}
where $e^{2\pi i c \epsilon_i(s,d)}$ are some $s,d$-dependent pure phases that are calculable (in fact they are always sixteenth roots of unity). 

Let us first for simplicity take the case $c\equiv 0 ~(\text{mod}~16)$. The sums in (\ref{eq:ThirdFourthv2}) have all $\epsilon_i$'s drop out and are simply Eisenstein series under $\Gamma_0(2)$. For $c \equiv 0 ~(\text{mod}~4), c>4$, 
\begin{align}
\sum_{\gamma \in \Gamma_\infty \backslash \Gamma_0(2)} (s\tau+d)^{-c/2} &= 1 + \frac{c}{(1-2^{\frac c2}) B_{c/2}} \sum_{k=1}^{\infty} \frac{k^{\frac c2-1}q^k(-1)^k}{1-q^k}
\label{eq:EisenGamma}
\end{align}
where $B_{n}$ is the $n^{\text{th}}$ Bernoulli number, defined as 
\be
\frac{x}{e^x-1} = \sum_{n=0}^{\infty} \frac{B_n x^n}{n!}.
\ee

Although for other values of $c$ mod 16, we do not have a closed form expression for the sum in (\ref{eq:ThirdFourthv2}), the sum converges for $c>4$. Therefore we can simply evaluate it numerically. We have checked numerically that adding (\ref{eq:ThirdFourthv2}), its complex conjugate, (\ref{eq:SecondFourv3}), and (\ref{eq:AlexEd}) does \emph{not} give the average partition function of a $\mathbb Z_2$ orbifold (\ref{eq:avgorbz2}). 

\section{K\"ahler moduli space} \label{app:Kahler}
In this appendix we derive the differential equation (\ref{mainDE}) satisfied by the Siegel-Narain theta function (\ref{thetacpx}), which we repeat here for convenience:
\be\label{mainDE_app}
\Big(\Delta_{\cH} - c\tau_2\frac{\partial}{\partial\tau_2}-  \Delta_{\cM_K} \Big)\Theta(m,\tau)= 0\ .
\ee
To do this, we first compute the Laplacian on the K\"ahler submanifold of the Narain moduli space $\cM_K$ in subsection \ref{app_lap}. We will then act with the Laplacian on the theta function in subsection \ref{app_lapact} and derive the differential equation.

\subsection{$\Delta_{\cM_K}$}\label{app_lap}
We consider $\sigma$-models with target spaces being toroidal orbifolds $T^{2d}/\bZ_N$, $N=3,4,6$.  As discussed in section \ref{avgzn}, the complex structure is fixed under the action of the cyclic group. The associated moduli space is then a K\"ahler submanifold of the Narain moduli space which we denote as ${\cal M}_K$. The coordinates on ${\cal M}_K$ are real $\frac c2\times \frac c2$ matrices $G_{i\bar j}$ and $B_{i\bar j}$. We would like to derive the Laplace operator on the K\"ahler subspace, $\Delta_{\cM_K}$.

We consider complex coordinates on the target space. The action of the world sheet $\sigma$-model is of the form:
\be\label{wsh_action}
S=\frac{1}{4\pi\alpha'}\int d^2\sigma\big[G_{i\bar\jmath}(\partial X^i\bar\partial X^{\bar\jmath}+\partial X^{\bar\jmath}\bar\partial X^i)+
iB_{i\bar\jmath}(\partial X^i\bar\partial X^{\bar\jmath}-\partial X^{\bar\jmath}\bar\partial X^i)\big]
\ee
where $G_{i\bar\jmath}$ is a Hermitian metric with $G_{ij}=G_{\bar\imath\bar\jmath}=0$. Likewise, for the $B$ fields $B_{ij}=B_{\bar\imath\bar\jmath}=0$. 
The exactly marginal operators of the theory are:
\be\label{O}
\cO=\delta G_{i\bar\jmath}(\partial X^i\bar\partial X^{\bar\jmath}+\partial X^{\bar\jmath}\bar\partial X^i)+
i\delta B_{i\bar\jmath}(\partial X^i\bar\partial X^{\bar\jmath}-\partial X^{\bar\jmath}\bar\partial X^i)
\ee
and the 2-point function gives the Zamolodchikov metric on the conformal manifold:
\be\label{MDmetricC}
\langle\cO(1,1)\;\cO(0,0)\rangle=G^{i\bar\ell}G^{\bar\jmath k}(\delta G_{i\bar\jmath}\delta G_{k\bar\ell}+\delta B_{i\bar\jmath}\delta B_{\bar\ell k})\ .
\ee
Thus, on ${\cM_K}$ we have
\be\label{MDmetricC_ii}
{\rm d}s^2=G^{i\bar\ell}G^{\bar\jmath k}({\rm d}G_{i\bar\jmath}{\rm d}G_{k\bar\ell}-{\rm d}B_{i\bar\jmath}{\rm d}B_{k\bar\ell})
\ee
and the metric $g_{\mu\nu}$ is block diagonal with the two blocks given by
\be\label{gijC}
\tilde g_{\mu\nu}=G^{i\bar\ell}G^{\bar\jmath k}\ ,\qquad \mu=(i,\bar\jmath)\;\;{\rm and}\;\;\nu=(k,\bar\ell)\ ,
\ee
for the $G$ coordinates and $-\tilde g_{\mu\nu}$ given for the $B$ coordinates.

Using the standard formula for the Laplace-Beltrami operator on curved space, the Laplacian corresponding to the first term on the \textsc{rhs} of eq. (\ref{MDmetricC_ii}), which we denote by $\Delta^G_{{\cal M}_K}$, is
\be\label{laplC}
\Delta^G_{{\cal M}_K}=\frac{-1}{\sqrt{|g|}}\partial_{G_{i\bar\jmath}}\Big(\sqrt{|g|}\,(G_{i\bar\ell}G_{\bar\jmath k})\,\partial_{G_{k\bar\ell}}\Big)\ .
\ee
Note that because of the block diagonal form the determinant satisfies $|g|=|\tilde g|^2$. This leads to
\be\label{laplC_ii}
\Delta^G_{{\cal M}_K}=-\bigg(\frac1{|\tilde g|}\Big(\frac{\partial|\tilde g|}{\partial{G_{i\bar\jmath}}}\Big)G_{i\bar\ell}G_{\bar\jmath k}\,\partial_{G_{k\bar\ell}}+
\frac{\partial(G_{i\bar\ell}G_{\bar\jmath k})}{\partial{G_{i\bar\jmath}}}\,\partial_{G_{k\bar\ell}}+
G_{i\bar\ell}G_{\bar\jmath k}\,\partial_{G_{i\bar\jmath}}\partial_{G_{k\bar\ell}}\bigg)\ .
\ee
The derivative of the determinant is given by
\bea\label{laplC_iii}
&&\frac{\partial|{\rm det}\tilde g_{\mu\nu}|}{\partial G_{i\bar\jmath}}=
|\tilde g|\,G_{i'\bar\ell'}G_{\bar\jmath' k'}\frac{\partial(G^{i'\bar\ell'}G^{\bar\jmath' k'})}{\partial G_{i\bar\jmath}}\\
&&\qquad\qquad\;=\frac{c|\tilde g|}2\,G_{i'\bar\ell'}\frac{\partial(G^{i'\bar\ell'}+G^{\bar\ell'i'})}{\partial G_{i\bar\jmath}}+
\frac{c|\tilde g|}2\,G_{\bar\jmath' k'}\frac{\partial(G^{\bar\jmath' k'}+G^{k'\bar\jmath'})}{\partial G_{i\bar\jmath}}=-c|\tilde g|G^{i\bar\jmath}\nonumber
\eea
where we used the fact that $G^{ij}=G^{\bar\imath\bar\jmath}=0$, as well as the following identities for the derivatives of matrices:
\be\label{matrix_der}
\frac{\partial G_{ij}}{\partial G_{kl}}=\delta_{~i}^k\delta_j^{~l}\ ,\quad\frac{\partial G^{ij}}{\partial G_{kl}}=-G^{ik}G^{lj}\ ,
\quad\frac{\partial{\rm det}g_{ij}}{\partial G_{kl}}=\Big(g^{ij}\frac{\partial g_{ij}}{\partial G_{kl}}\Big){\rm det}(g_{ij})\ .\quad
\ee

Inserting eq. (\ref{laplC_iii}) back in eq. (\ref{laplC_ii}) we find
\bea\label{laplC_iv}
&&\Delta^G_{{\cal M}_K}=-\bigg(\frac1{|\tilde g|}(-c|\tilde g|G^{i\bar\jmath})G_{i\bar\ell}G_{\bar\jmath k}\,\partial_{G_{k\bar\ell}}+\\
&&\qquad\qquad\quad+\Big(\delta^i_{~i}\delta_{\bar\ell}^{~\bar\jmath}G_{\bar\jmath k}+
G_{i\bar\ell}\delta^i_{~k}\delta_{\bar\jmath}^{~\bar\jmath}\Big)\partial_{G_{k\bar\ell}}
+G_{i\bar\ell}G_{\bar\jmath k}\,\partial_{G_{i\bar\jmath}}\partial_{G_{k\bar\ell}}\bigg)\nonumber\\
&&\qquad\quad= c G_{k\bar\ell}\partial_{G_{k\bar\ell}}-cG_{k\bar\ell}\partial_{G_{k\bar\ell}}
-G_{i\bar\ell}G_{\bar\jmath k}\,\partial_{G_{i\bar\jmath}}\partial_{G_{k\bar\ell}}
=-G_{i\bar\ell}G_{\bar\jmath k}\,\partial_{G_{i\bar\jmath}}\partial_{G_{k\bar\ell}}\ .\nonumber
\eea
The Laplacian corresponding to the second term on the \textsc{RHS} of eq. (\ref{MDmetricC_ii}), which we denote by $\Delta_{\cM_K}^B$, is much easier to compute since the metric is independent of $B$ fields:
\be\label{laplCB}
\Delta_{{\cal M}_K}^B=G_{i\bar\ell}G_{\bar\jmath k}\,\partial_{B_{i\bar\jmath}}\partial_{B_{k\bar\ell}}\ .
\ee
Putting together eqs. (\ref{laplC_iv}) and (\ref{laplCB}), we find the Laplacian on the K\"ahler submanifold:
\be\label{laplC_v}
\Delta_{{\cal M}_K}=-G_{i\bar\ell}G_{\bar\jmath k}(\partial_{G_{i\bar\jmath}}\partial_{G_{k\bar\ell}}-
\partial_{B_{i\bar\jmath}}\partial_{B_{k\bar\ell}})\ .
\ee

\subsection{$\Delta_{{\cal M}_K}\Theta$}\label{app_lapact}
We next apply the Laplacian $\Delta_{{\cal M}_K}$ on the lattice sum $\Theta$ (\ref{thetacpx}). Let us first consider the action of the first term on the \textsc{rhs} of eq. (\ref{laplC_v}) on each term $Q$  (\ref{latsumQ}) in the lattice sum. The first derivative $ \partial_{G_{k\bar\ell}}$ gives
	\be
	-\frac{2\pi\tau_2}{\alpha'}(w^k\bar w^{\bar\ell} - G^{\bar sk}G^{r\bar\ell}\bar v_{\bar s}v_r)Q\ .
	\ee
Taking the second derivative $\partial_{G_{i\bar\jmath}}$ contracted with $-G_{i\bar\ell}G_{\bar\jmath k}$ we obtain
	\begin{align}\label{DelQG}
	&-\Big(\frac{2\pi\tau_2}{\alpha'}\Big)^2(w^k\bar w^{\bar\ell} - \bar v^k v^{\bar\ell})(w^i\bar w^{\bar\jmath} - \bar v^{i}v^{\bar\jmath})G_{k\bar\jmath}G_{i\bar\ell}
	+\frac{2\pi\tau_2}{\alpha'}(G^{\bar si}G^{k\bar\jmath}G^{r\bar\ell}+G^{k\bar s}G^{r\bar\jmath}G^{i\bar\ell})\bar v_{\bar s}v_rG_{k\bar\jmath}G_{i\bar\ell}\nn\\
	&= -\Big(\frac{\pi\tau_2}{\alpha'}\Big)^2 (w_{\bar\jmath}\bar w_{i} - \bar v_{\bar\jmath}v_{i})(w^i\bar w^{\bar\jmath} - \bar v^{i}v^{\bar\jmath})
	+ \frac{\pi\tau_2}{\alpha'} c|v|^2\nn\\
	&= -\Big(\frac{\pi\tau_2}{\alpha'}\Big)^2\Big((|v|^2)^2 - 2(w\cdot v)(\bar w\cdot\bar v) + (|w|^2)^2\Big)
	+ \frac{\pi\tau_2}{\alpha'} c|v|^2
	\end{align}

For the second term on the \textsc{rhs} of eq. (\ref{laplC_v}) we use $\partial_{B_{k\bar\ell}} v_r = \delta_r^k\bar w^{\bar\ell}$ and $\partial_{B_{k\bar\ell}} \bar v_{\bar s} = - \delta_{\bar s}^{\bar\ell}w^{k}$. The first $\partial_{B_{k\bar\ell}}$ derivative gives
	\be
	-\frac{2\pi\tau_2}{\alpha'}(G^{k\bar s}\bar w^{\bar\ell}\bar v_{\bar s}-G^{r\bar\ell} v_rw^k )
	= -\frac{\pi\tau_2}{\alpha'} (\bar w^{\bar n}\bar v^{m}-v^{\bar n}w^m)\ .
	\ee
	Acting with the second derivative $\partial_{B_{i\bar\jmath}}$ we find
	\be
	\Big(\frac{\pi\tau_2}{\alpha'}\Big)^2 (\bar w^{\bar\ell}\bar v^{k}-v^{\bar\ell}w^k)(\bar w^{\bar\jmath}\bar v^{i}-v^{\bar\jmath}w^i) + 
	\frac{\pi\tau_2}{\alpha'} (G^{k\bar\jmath} \bar w^{\bar\ell}w^q + G^{\bar\ell i}w^k \bar w^{\bar\jmath})
	\ee
	which, after contraction $-G_{i\bar\ell}G_{\bar\jmath k}$, gives
	\be\label{DelQB}
	-\Big(\frac{\pi\tau_2}{\alpha'}\Big)^2 (2|w|^2|v|^2-(w\cdot v)^2-(\bar w\cdot \bar v)^2)+\frac{\pi\tau_2}{\alpha'}c |w|^2\ .
	\ee
Putting together eqs. (\ref{DelQG}) and (\ref{DelQB}) we find
	\be\label{DelQ}
	\Delta_{{\cal M}_K}Q =\\ \left(
	-\Big(\frac{\pi\tau_2}{\alpha'}\Big)^2\Big((|v|^2+|w|^2)^2-4(w\cdot v+\bar w\cdot \bar v)^2\Big)+\frac{\pi\tau_2}{\alpha'}c(|v|^2+|w|^2)\right)Q
	\ee

We next recall that the moduli space of the world sheet Riemann surface, $\Sigma$, is the upper half plane $\cal H$ with metric
\be\label{Hmetric}
ds^2=\tf1{\tau_2^2}(d\tau_1^2+d\tau_2^2)
\ee
and Laplacian
\be\label{Hlapl}
\Delta_{\cal H}=-{\tau_2^2}\Big(\frac{\partial^2}{\partial\tau_1^2}+\frac{\partial^2}{\partial\tau_2^2}\Big)\ .
\ee
The action of this Laplacian on $Q$ gives: 
	\be\label{DelH}
	\Delta_{\mathcal{H}}Q =-\tau_2^2\left( (\frac\pi{\alpha'})^2 (|v|^2+|w|^2)^2 -4\pi^2 (n\cdot  w+ \bar n \cdot \bar w)^2 \right)Q\ .
	\ee
Moreover, we have the identity:
	\be\label{Dtau2}
	c\tau_2\partial_{\tau_2}Q = -c\frac{\pi\tau_2}{\alpha'}(v^2+w^2)Q\ .
	\ee
All in all, using eqs. (\ref{DelQ}), (\ref{DelH}), and (\ref{Dtau2}), and using  $w\cdot v = \alpha'( n\cdot w)$, we derive the differential equation (\ref{mainDE_app}).

\section{$\mathcal{N}=4$ modular kernels}
\label{sec:c6kernels}

In this appendix we will work out the modular kernels for the small $\mathcal{N}=4$ superconformal algebra at central charge $6$. The $S$ and $T$ kernels were computed in \cite{Eguchi:1987wf, Eguchi:1988af}. The $S$ kernels are:
\begin{align}
\chi^G\(-\frac1\tau, \frac z\tau\) &= -e^{2\pi i \frac{z^2}\tau}\(2 \chi^M(\tau, z) + 2 \int_0^{\infty} dp' \sinh(\pi p') \tanh(\pi p') \chi^{p'}(\tau, z)\) \nn\\
&= -e^{2\pi i \frac{z^2}\tau}\(2 \chi^M(\tau, z) + \sqrt 2 \int_{\frac18}^{\infty} dh \frac{\sinh^2\(\sqrt 2\pi \sqrt{h-\frac18}\) }{\cosh\(\sqrt 2\pi \sqrt{h-\frac18}\)\sqrt{h- \frac18}} (\chi^{G}(\tau, z)+2\chi^{M}(\tau,z))q^h\) \nn\\
\chi^M\(-\frac1\tau, \frac z\tau\) &= e^{2\pi i \frac{z^2}\tau}\(\chi^M(\tau, z) - \int_0^{\infty}  \frac{dp'}{\cosh(\pi p')} \chi^{p'}(\tau, z)\) \nn\\
&= e^{2\pi i \frac{z^2}\tau}\(\chi^M(\tau, z) - \frac1{\sqrt 2} \int_{\frac18}^{\infty}  \frac{dh}{\cosh\(\sqrt 2\pi \sqrt{h-\frac18}\)\sqrt{h- \frac18}} (\chi^{G}(\tau, z)+2\chi^{M}(\tau,z))q^h\) \nn\\
\chi^{p}\(-\frac1\tau, \frac{z}{\tau}\) &= -2e^{2\pi i \frac{z^2}{\tau}}\int_0^\infty dp' \cos(2\pi p p') \chi^{p'}(\tau, z) \nn\\
\Bigg[\chi^{G}(-\frac1\tau, \frac z\tau)+&2\chi^{M}(-\frac1\tau, \frac z\tau)\Bigg]e^{-\frac{2\pi i}\tau h} = -\sqrt2 e^{2\pi i \frac{z^2}\tau} \int_{\frac18}^\infty \frac{dh'\cos\(4\pi\sqrt{\(h-\frac18\)\(h'-\frac18\)}\)}{\sqrt{h-\frac18}}  (\chi^{G}(\tau, z)+2\chi^{M}(\tau,z))q^{h'}.
\label{eq:Skernels}
\end{align}
The $T$ kernels are more straightforward: 
\begin{align}
\chi^G\(\tau+1,z\) &= \chi^G\(\tau,z\) \nn\\
\chi^M\(\tau+1,z\) &= \chi^M\(\tau,z\) \nn\\
\chi^{p}\(\tau+1,z\) &= e^{2\pi i \(\frac{p^2}2 + \frac18\)} \chi^{p}(\tau, z) = e^{2\pi i h} \left[(\chi^{G}(\tau, z)+2\chi^{M}(\tau,z)\right]q^{h}.
\label{eq:Tkernels}
\end{align}
For convenience we have written the kernels in both ``Liouville notation," as well as more standard notation. 

We will now work out the $ST^nS$ kernels for integer $n$. Our strategy will follow that of Appendix D of \cite{Benjamin:2019stq}. Let us first do the $S$ transform:
\begin{align}
\chi^G\(ST^nS\tau, ST^nSz\)&=-e^{2\pi i \frac{z^2}{\tau(n\tau-1)}}\(2\chi^M(T^n S\tau, T^n S z) + 2\int_0^\infty dp' \sinh\(\pi p'\) \tanh\(\pi p'\) \chi^{p'}(T^nS\tau, T^nSz)\)
\label{eq:firststepsts}
\end{align}
where $\gamma\tau = \frac{a\tau+b}{c\tau+d}$, and $\gamma z = \frac{z}{c\tau+d}$. Since $\chi^M$ is invariant under $T$, we can remove the $T^n$ for $\chi^M$ on the RHS of (\ref{eq:firststepsts}), and for $\chi^{p'}$ we pick up a phase:
\begin{align}
\chi^G\(ST^nS\tau, ST^nSz\)&=-e^{2\pi i \frac{z^2}{\tau(n\tau-1)}}\(2\chi^M(S\tau, S z) + 2e^{\frac{2\pi i n}8}\int_0^\infty dp' \sinh\(\pi p'\) \tanh\(\pi p'\) e^{\frac{2\pi i n (p')^2}2} \chi^{p'}(S\tau, Sz)\).
\label{eq:secstepsts}
\end{align}
Finally we can do the $S$ transform which gives:
\begin{align}
\chi^G\(ST^nS\tau, ST^nSz\)&=-e^{2\pi i \frac{z^2}{\tau(n\tau-1)}}\(2\chi^M(S\tau, S z) + 2e^{\frac{2\pi i n}8}\int_0^\infty dp' \sinh\(\pi p'\) \tanh\(\pi p'\) e^{\frac{2\pi i n (p')^2}2} \chi^{p'}(S\tau, Sz)\) \nn\\
&= -e^{2\pi i \frac{nz^2}{n\tau-1}}\Bigg[2\chi^M(\tau, z)  - 2\int_0^{\infty}  \frac{dp}{\cosh(\pi p)} \chi^{p}(\tau, z) \nn\\
&~~~~~- 4 e^{\frac{2\pi i n}8} \int_0^{\infty} dp' \sinh\(\pi p'\) \tanh\(\pi p'\) e^{\frac{2\pi i n (p')^2}2} \int_0^\infty dp \cos\(2\pi p p'\) \chi^p(\tau, z)\Bigg] \nn\\
&= e^{2\pi i \frac{nz^2}{n\tau-1}}\Bigg[-2\chi^M(\tau, z)  + 2\int_0^{\infty}  \frac{dp}{\cosh(\pi p)} \chi^{p}(\tau, z) \nn\\
&~~~~~+ 2 e^{\frac{2\pi i n}8} \int_{-\infty}^{\infty} dp' \sinh\(\pi p'\) \tanh\(\pi p'\) e^{\frac{2\pi i n (p')^2}2} \int_{0}^\infty dp  \cos\(2\pi p p'\) \chi^p(\tau, z)\Bigg] \nn\\
&\equiv e^{2\pi i \frac{nz^2}{n\tau-1}}\Bigg[-2\chi^M(\tau, z)  + \int_0^\infty  dp K^{ST^nS}(p) \chi^p(\tau, z) \Bigg]. 
\label{eq:thirdstepsts}
\end{align}
To read off the kernel $K^{ST^nS}(p)$, we would like to do the integral:
\begin{align}
\int_{-\infty}^{\infty} dp' \sinh(\pi p') &\tanh(\pi p') e^{\pi i n (p')^2} e^{2\pi i p p'} = \frac14 \int_{-\infty}^{\infty} dp' \frac{\(e^{\pi p'} - e^{-\pi p'}\)^2}{\cosh(\pi p')} e^{\pi i n (p')^2} e^{2\pi i p p'} \nn\\
&= \frac14 \int_{-\infty}^{\infty} dp' \frac{e^{2\pi(1+ i p) p'}}{\cosh(\pi p')} e^{\pi i n (p')^2} + \frac14 \int_{-\infty}^{\infty} dp' \frac{e^{2\pi(-1+ i p) p'}}{\cosh(\pi p')} e^{\pi i n(p')^2} -\frac12 \int_0^{\infty} dp' \frac{e^{2\pi i p p'}}{\cosh(\pi p')} e^{\pi i n (p')^2}
\label{eq:fourthstepsts}
\end{align}
The last integral in (\ref{eq:fourthstepsts}) converges. More generally the integral
\be
\int_{-\infty}^\infty dp' \frac{e^{\pi i n (p')^2 + 2\pi z p'}}{\cosh(\pi p')}
\label{eq:mordell}
\ee
converges if $\text{Im}(n) > 0$, or if $n \in \mathbb R, |\text{Re}(z)| < \frac12$. For example, the first integral in (\ref{eq:fourthstepsts}) converges if $\frac 12 < \text{Im}(p) < \frac 32$, the second converges if $-\frac32 < \text{Im}(p) < -\frac12$, and the third if $-\frac12 < \text{Im}(p) < \frac12$. Alternatively we could give $n$ a small (positive) imaginary part and all three would converge.

Remarkably, L. J. Mordell considered precisely the integral in (\ref{eq:mordell}) in 1933 \cite{Mordell}. It is now known as a Mordell integral, and is closely related to the theory of mock modular forms. Following \cite{Zwegers:2008zna}, let us denote the function $h(\tau, z)$ as
\be
h(\tau, z) = \int_{-\infty}^\infty dp'  \frac{e^{\pi i \tau (p')^2 - 2\pi z p'}}{\cosh(\pi p')}.
\ee
Our kernel is given by:
\begin{align}
&K^{ST^nS}(p) = \frac2{\cosh(\pi p)} \nn\\ 
&~~+ \frac{e^{\frac{2\pi i n}8}}4\(h(n, 1+ip) + h(n,1-ip) + h(n,-1+ip) + h(n,-1-ip)-2h(n, ip)-2h(n,-ip)\).
\label{eq:kernel}
\end{align}

We can now use properties of the function $h(\tau, z)$. In particular, using the following properties found in \cite{Zwegers:2008zna}:
\begin{align}
h(\tau, z) &= h(\tau, -z) \nn\\
h(\tau, z) + h(\tau, z+1) &= \frac{2}{\sqrt{-i\tau}} e^{\frac{i \pi (z+\frac12)^2}\tau},
\label{eq:HProp}
\end{align}
we can rewrite (\ref{eq:kernel}) as
\be
K^{ST^nS}(p) = \frac{2}{\cosh(\pi p)} + \frac{e^{\frac{2\pi i (n+1)}8}}{\sqrt n}\left[ e^{\frac{i\pi}n\(ip+\frac12\)^2} + e^{\frac{i\pi}n\(ip-\frac12\)^2} \right] - 2 e^{\frac{2\pi i n}8} h(n, ip)
\ee
or equivalently 
\be
K^{ST^nS}(p) = \frac{2}{\cosh(\pi p)} + \frac{2 \cosh\(\frac{\pi p}n\)}{\sqrt{n}} e^{-\frac{2\pi i}n\(\frac{p^2}2+\frac18\)} e^{2\pi i \(\frac n8 + \frac18 + \frac1{4n}\)} - 2 e^{\frac{2\pi i n}8} h(n, ip).
\label{eq:kernelstep4orso}
\ee
By plugging in $\tau=0$ in Property $(6)$ in Proposition $1.2$ of \cite{Zwegers:2008zna}, we find that
\be
h(1, ip) = \text{sech}(\pi p)\left[ e^{2\pi i\(\frac78\)} + i e^{-ip^2\pi}\right].
\label{eq:H1exact}
\ee
Plugging (\ref{eq:H1exact}) into (\ref{eq:kernelstep4orso}) for $n=1$ gives
\be
K^{STS}(p) =  -2\sinh(\pi p)\tanh(\pi p) e^{-2\pi i \(\frac{p^2}2+\frac18\)}
\ee
which is precisely what we expect from (\ref{eq:Skernels}) under the $T^{-1}S T^{-1}$ transformation. If we knew the general expression for $h(n, ip)$ for arbitrary positive integer $n$, then we would get the full modular kernel $K^{ST^nS}(p)$. Interestingly it seems there is no known analytic expression for $h(n, ip)$ for $n>1$. However, we can numerically evaluate the integral to extremely high precision. For example, from numerically evaluating $h(2,ip)$ to extremely high precision, we conjecture that $h(2,ip)$ takes the following exact form:
\begin{align}
h(2,ip) &= e^{-\frac{2\pi i}4} \text{sech}(\pi p) + \sqrt{2}e^{-\frac{2\pi i}2\(\frac{p^2}2-\frac38\)}\(\frac12\text{sech}\(\frac{\pi p}2\) + \sum_{n=1}^{\infty} a(n) \text{sech}\(\frac{(2n+1)\pi p}{2}\)\),
\end{align}
where $a(n)$ are integers, with the first $90$ values given by:
\begin{align}
a(n) &= 2, ~~ n = 16, 28, 32, 37, 49, 64, 72, 85, 88, \nn\\
a(n) &= 1, ~~ n = 1, 4, 5, 9, 12, 13, 21, 29, 33, 40, 41, 53, 60, 65, 69, 81, 84, 89, \nn\\
a(n) &= 0, ~~ n = 3, 8, 10, 11, 15, 17, 20, 23, 24, 25, 31, 34, 35, 36, 38, 39, 42, 44, 45, \nn\\
&~~~~~~~~~~~~~~~ 46, 48, 51, 52, 56, 57, 59, 61, 63, 66, 68, 70, 73, 75, 76, 77, 80
,83, 87, \nn\\
a(n) &= -1, ~~ n = 2, 6, 14, 18, 26, 30, 50, 54, 62, 74, 78, 86, 90, \nn\\
a(n) &= -2, ~~n = 7, 19, 22, 27, 43, 47, 55, 58, 67, 71, 79, \nn\\
a(n) &= -4, ~~ n = 82.
\end{align}

Moreover, for $h(2,ip)$, we find that in the limit of large and small $p$ respectively, we have: 
\begin{align}
h(2,ip) &\sim \sqrt{2}e^{-\frac{\pi p}2}, ~~~p~\text{large} \nn\\
h(2,ip) &= \(\sqrt{2} e^{2\pi i\(\frac{3}{16}\)}-i\) + \(\frac{\pi(4\cos\(\frac\pi8\) - 3\pi\sin\(\frac\pi8\))}{4\sqrt 2} + \(\frac{\pi^2}2 - \frac{3\pi^2 \cos(\frac \pi8)}{4\sqrt2} - \frac{\pi \sin\(\frac\pi8\)}{\sqrt2}\)i\)p^2 + \mc{O}(p^4).
\label{eq:h20limits}
\end{align}

Plugging in the second line of (\ref{eq:h20limits}) into (\ref{eq:kernelstep4orso}) at $n=2$ gives the kernel in the small $p$ limit. In particular, it will become
\be
K^{ST^2S}(p) = \sqrt{2} e^{-i \pi \(\frac{p^2}2 + \frac18\)}\(1 - \frac{7\pi^2}{8}p^2 +  \frac{113\pi^4}{384} p^4 + \mathcal{O}(p^6)\),
\label{eq:ST2Sc6K}
\ee
where every coefficient in the series in (\ref{eq:ST2Sc6K}) is a real number.

\section{Extended $\mathcal{N}=2$ modular kernels}
\label{app:mwkcy}

In this appendix we will compute the Poincar\'e sum of the vacuum character of the extended $\mathcal{N}=2$ algebra. We will show that the density of states is not positive definite for $c>3$.

We first calculate the kernels of the extended $\mathcal{N}=2$ algebra at $c=3\hat c$. There will in general be $\hat c-1$ long representations and $\hat c$ short representations. The kernels can be found in \cite{Eguchi:2003ik}. In Liouville notation, they are given by the following. For $\hat c$ even:
\begin{align}
\chi_G^{\hat c}\(-\frac1\tau, \frac z \tau\) &= e^{-i\pi \frac{\hat c}2} e^{\frac{i\pi \hat c z^2}{\tau}}\Bigg[\frac{1}{\sqrt{\hat c-1}}\sum_{Q'=-\frac{\hat c}2+1}^{\frac{\hat c}2 -1} \int_0^\infty dp' \frac{\sinh(\pi\sqrt{\hat c-1}p')\sinh(2\pi \frac{p'}{\sqrt{\hat c- 1}})}{|\cosh(\pi(\frac{p'}{\sqrt{\hat c-1}}+i \frac{Q'}{\hat c-1}))|^2}\chi(\tau, z)^{p'}_{Q'} \nn\\
&+ 2\chi_M(\tau, z)_{\hat j = \frac{\hat c}2}\Bigg]
\label{eq:ChiGEven}
\end{align}
and for $\hat c$ odd:
\begin{align}
\chi_G^{\hat c}\(-\frac1\tau, \frac z \tau\) &= e^{\frac{i\pi \hat c z^2}{\tau}}e^{-\frac{i\pi \hat c}2}\Bigg[\frac{1}{\sqrt{\hat c-1}}\sum_{Q'=-\frac{\hat c-3}{2}}^{\frac{\hat c-1}2} \int_0^\infty dp' \frac{\sinh(\pi\sqrt{\hat c-1}p')\sinh(2\pi \frac{p'}{\sqrt{\hat c- 1}})}{|\cosh(\pi(\frac{p'}{\sqrt{\hat c-1}}+i \frac{Q'}{\hat c-1}))|^2}\chi(\tau, z)^{p'}_{Q'}\Bigg]
\label{eq:ChiGOdd}
\end{align}
The long characters transform more simply
\begin{align}
\chi\(-\frac1\tau, \frac z\tau\)^{p}_Q &= e^{-i\pi \frac{\hat c}2} e^{\frac{i\pi \hat c z^2}\tau}\frac{2}{\sqrt{\hat c-1}}\int_0^{\infty} dp' \cos(2\pi p p')\sum_{Q'=0}^{\hat c-2} e^{-2\pi i \frac{QQ'}{\hat c-1}} \chi(\tau, z)^{p'}_{Q'}.
\label{eq:AllParLong}
\end{align}
Finally for $\hat c$ even we will need the $S$ transform of the BPS character $\chi_M$ with charge $Q=\frac{\hat c}2$.
\begin{align}
\chi_M\(-\frac1\tau, \frac z\tau\)_{Q=\frac{\hat c} 2} &= e^{\frac{i \pi \hat c z^2}{\tau}}\Bigg[\frac{(-1)^{\frac{\hat c}2}}{2\sqrt{\hat c -1}}\sum_{Q'=-\frac{\hat c}2+1}^{\frac{\hat c}2 -1} e^{-\frac{\pi i \hat c Q'}{\hat c-1}}\int_0^{\infty} dp'  \frac{\cosh(\frac{\pi p'}{\sqrt{c-1}})(1+e^{\frac{2\pi i Q'}{\hat c-1}})}{|\cosh(\pi(\frac{p'}{\sqrt{\hat c-1}}+i\frac{Q'}{\hat c-1}))|^2}\chi(\tau, z)_{Q'}^{p'}\nn\\
&+\chi_M(\tau, z)_{Q=\frac{\hat c}2}\Bigg].
\label{eq:ChiMEven}
\end{align}

\subsection{Even $\hat c$}
Let us now compute the $ST^nS$ kernel for these characters. First we will do the case $\hat c$ is even:
\begin{align}
\chi_G(ST^nS &\tau, ST^nS z) = (-1)^{\frac{\hat c}2}e^{2\pi i \frac{z^2}{\tau(n\tau-1)}}\Big(2\chi^M(T^n S\tau, T^n S z)_{Q=\frac{\hat c}2} \nn\\
&~~~+ \frac1{\sqrt{\hat c-1}}\sum_{Q'=-\frac{\hat c}2+1}^{\frac{\hat c}2-1}\int_0^\infty dp'\frac{\sinh(\pi\sqrt{\hat c-1}p')\sinh(2\pi \frac{p'}{\sqrt{\hat c- 1}})}{|\cosh(\pi(\frac{p'}{\sqrt{\hat c-1}}+i \frac{Q'}{\hat c-1}))|^2} \chi^{p'}(T^nS\tau, T^nSz)_{Q=Q'}\Big) \nn\\
&=  (-1)^{\frac{\hat c}2}e^{2\pi i \frac{z^2}{\tau(n\tau-1)}}\Big(2\chi^M(S\tau, S z)_{Q=\frac{\hat c}2} \nn\\
&~~~+ \frac{e^{\frac{2\pi i n(\hat c-1)}{8}}}{\sqrt{\hat c-1}}\sum_{Q'=-\frac{\hat c}2+1}^{\frac{\hat c}2-1}e^{\frac{2\pi i n (Q')^2}{2(\hat c-1)}-\frac{2\pi i n Q'}2}\int_0^\infty dp'\frac{\sinh(\pi\sqrt{\hat c-1}p')\sinh(2\pi \frac{p'}{\sqrt{\hat c- 1}})}{|\cosh(\pi(\frac{p'}{\sqrt{\hat c-1}}+i \frac{Q'}{\hat c-1}))|^2} e^{\frac{2\pi i n (p')^2}{2}} \chi^{p'}(S\tau, Sz)_{Q'}\Big) \nn\\
&= e^{2\pi i \frac{n z^2}{n\tau-1}} \Bigg(2(-1)^{\frac{\hat c}2}\chi_M(\tau, z)_{\frac{\hat c}2} + \frac{1}{\sqrt{\hat c -1}}\sum_{Q'=-\frac{\hat c}2+1}^{\frac{\hat c}2 -1} e^{-\frac{\pi i \hat c Q'}{\hat c-1}}\int_0^{\infty} dp'  \frac{\cosh(\frac{\pi p'}{\sqrt{\hat c-1}})(1+e^{\frac{2\pi i Q'}{\hat c-1}})}{|\cosh(\pi(\frac{p'}{\sqrt{\hat c-1}}+i\frac{Q'}{\hat c-1}))|^2}\chi(\tau, z)_{Q'}^{p'}\nn\\
&~~+\frac{2e^{\frac{2\pi i n(\hat c-1)}{8}}}{\hat c-1}\sum_{Q'=-\frac{\hat c}2+1}^{\frac{\hat c}2-1}e^{\frac{2\pi i n (Q')^2}{2(\hat c-1)}-\frac{2\pi i n Q'}2}\int_0^\infty dp'\frac{\sinh(\pi\sqrt{\hat c-1}p')\sinh(2\pi \frac{p'}{\sqrt{\hat c- 1}})}{|\cosh(\pi(\frac{p'}{\sqrt{\hat c-1}}+i \frac{Q'}{\hat c-1}))|^2} e^{\frac{2\pi i n (p')^2}{2}} \nn\\
&~~~~~~~~~~~~~~~~~~~~~~~~~~\times \int_0^\infty dp'' \cos(2\pi p' p'')\sum_{Q''=0}^{\hat c-2} e^{-2\pi i \frac{Q' Q''}{\hat c-1}}\chi^{p''}(\tau, z)_{Q''}\Bigg).
\label{eq:lotsofmanipulations}
\end{align}
The contribution to, say, the $Q=0$ long multiplet is then given by
\begin{align}
\frac{2}{\sqrt{\hat c-1}} &\text{sech}\(\frac{\pi p}{\sqrt{\hat c-1}}\) \nn\\
&+\frac{2e^{\frac{2\pi i n(\hat c-1)}{8}}}{\hat c-1}\sum_{Q'=-\frac{\hat c}2+1}^{\frac{\hat c}2-1}e^{\frac{2\pi i n (Q')^2}{2(\hat c-1)}-\frac{2\pi i n Q'}2}\int_0^\infty dp'\frac{\sinh(\pi\sqrt{\hat c-1}p')\sinh(2\pi \frac{p'}{\sqrt{\hat c- 1}})}{|\cosh(\pi(\frac{p'}{\sqrt{\hat c-1}}+i \frac{Q'}{\hat c-1}))|^2} e^{\frac{2\pi i n (p')^2}{2}}\cos(2\pi p p') 
\label{eq:Evencontr}
\end{align}
We can give $n$ a small imaginary part, i.e. set $n=2+i \epsilon$ so that the integral converges, and then take the $\epsilon\rightarrow0$ limit. It appears numerically that (\ref{eq:Evencontr}) is nonzero in the $p\rightarrow0$ limit. Assuming that (\ref{eq:Evencontr}) then has a term that goes as $e^{-i \pi \(\frac{p^2}2+\frac18\)}$, this implies the Poincar\'e sum is negative, because the $K^{ST^2S}(h) K^{ST^2S}(\bar h)$ kernel scales as $\mathcal{O}(1)$, whereas the $K^S(h) K^S(\bar h)$ scales as $\mathcal{O}(p^2)$. It would be good to more rigorously show this as we did for $\hat c =2$. 

\subsection{Odd $\hat c$}
Now let us do the case of $\hat c$ odd. 
\begin{align}
\chi_G(ST^nS &\tau, ST^nS z) = \frac{e^{2\pi i \frac{z^2}{\tau(n\tau-1)}}e^{-\frac{i\pi \hat c}2}}{\sqrt{\hat c -1}}\sum_{Q'=-\frac{\hat c-3}2}^{\frac{\hat c-1}2}\int_0^\infty dp'\frac{\sinh(\pi\sqrt{\hat c-1}p')\sinh(2\pi \frac{p'}{\sqrt{\hat c- 1}})}{|\cosh(\pi(\frac{p'}{\sqrt{\hat c-1}}+i \frac{Q'}{\hat c-1}))|^2} \chi^{p'}(T^nS\tau, T^nSz)_{Q'} \nn\\
&= \frac{e^{2\pi i \frac{z^2}{\tau(n\tau-1)}}e^{-\frac{i\pi \hat c}2}e^{\frac{2\pi i n(\hat c-1)}{8}}}{\sqrt{\hat c-1}}\Big(\sum_{Q'=-\frac{\hat c-3}2}^{\frac{\hat c-1}2} e^{\frac{2\pi i n (Q')^2}{2(\hat c-1)}-\frac{2\pi i n Q'}2} \nn\\
&~~~~~~~~~~~~~~~~~~~~~~~~~~~~~~~~~~~~~~~~~~~~~\times \int_0^\infty dp' \frac{\sinh(\pi\sqrt{\hat c-1}p')\sinh(2\pi \frac{p'}{\sqrt{\hat c- 1}})}{|\cosh(\pi(\frac{p'}{\sqrt{\hat c-1}}+i \frac{Q'}{\hat c-1}))|^2} e^{\frac{2\pi i n (p')^2}{2}} \chi^{p'}(S\tau, Sz)_{Q'}\Big) \nn\\
&=-\frac{2e^{2\pi i \frac{n z^2}{n\tau-1}}e^{\frac{2\pi i n(\hat c-1)}{8}}}{\hat c-1}\Bigg(\sum_{Q'=-\frac{\hat c-3}2}^{\frac{\hat c-1}2} e^{\frac{2\pi i n (Q')^2}{2(\hat c-1)}-\frac{2\pi i n Q'}2} \nn\\
&~~~\times \int_0^\infty dp' \frac{\sinh(\pi\sqrt{\hat c-1}p')\sinh(2\pi \frac{p'}{\sqrt{\hat c- 1}})}{|\cosh(\pi(\frac{p'}{\sqrt{\hat c-1}}+i \frac{Q'}{\hat c-1}))|^2} e^{\frac{2\pi i n (p')^2}{2}} \int_0^{\infty} dp'' \cos(2\pi p' p'') \sum_{Q''=0}^{\hat c-2} e^{-2\pi i \frac{Q'Q''}{\hat c-1}} \chi(\tau, z)_{Q''}^{p''}\Bigg).
\label{eq:OddAS}
\end{align}
The contribution to the $Q=0$ long multiplets is then given by
\begin{align}
-\frac{2e^{\frac{2\pi i n(\hat c-1)}{8}}}{\hat c-1}\sum_{Q'=-\frac{\hat c-3}{2}}^{\frac{\hat c-1}2} e^{\frac{2\pi i n (Q')^2}{2(\hat c-1)}-\frac{2\pi i nQ'}2} \int_0^\infty dp'\frac{\sinh(\pi\sqrt{\hat c-1}p')\sinh(2\pi \frac{p'}{\sqrt{\hat c- 1}})}{|\cosh(\pi(\frac{p'}{\sqrt{\hat c-1}}+i \frac{Q'}{\hat c-1}))|^2} e^{\frac{2\pi i n (p')^2}{2}} \cos(2\pi pp')
\label{eq:Oddasdf}
\end{align}
Again, we can set $n=2+i\epsilon$ for small $\epsilon$ and numerically see that (\ref{eq:Oddasdf}) is nonzero as $p\rightarrow0$. 

\subsection{Special case: $\hat c=3$}

Let us consider the special case of $\hat c=3$, which is relevant for CFTs with target space Calabi-Yau threefold. We have already seen that the Poincar\'e sum of the vacuum character is not positive definite. What about the addition of the other half- and quarter-BPS states? Remarkably a generic CY3 (with no enhanced symmetry) has \emph{no} quarter-BPS states! In other words, all BPS states are generically given by
\be
Z^{BPS} = \chi^0 \overline{\chi^0} + h^{1,1}\(\chi^1 \overline{\chi^1} + \chi^{-1} \overline{\chi^{-1}}\)+ h^{2,1}\(\chi^1 \overline{\chi^{-1}} + \chi^{-1} \overline{\chi^{1}}\)
\label{eq:ZBPSCy3}
\ee
We have argued that the Poincar\'e sum of the first term,  $\chi^0 \overline{\chi^0}$ does not lead to a positive definite density of states. Would the addition of the two other terms cure the negativity in the odd spin, low twist states? The answer is \emph{no}. Unlike in the case of K3, there are only a finite number of terms in the sum (\ref{eq:ZBPSCy3}). We can then do the modular sum of each individually. 

Let us explicit compute these terms. Remarkably, the formulas in the previous section simplify substantially in the case of $\hat c =3$. In particular, if we take (\ref{eq:OddAS}) and plug in $\hat c= 3$, we get:
\begin{align}
\chi_G(ST^nS\tau, ~&ST^n Sz) = \frac{e^{\frac{2\pi i n z^2}{n\tau-1}}}{\sqrt{-in}} \int_0^\infty dp~e^{-\frac{i\pi p^2}n}\Bigg[\(-1+e^{\frac{in \pi}2} - e^{\frac{i\pi}{2n}}\(1+e^{\frac{in\pi}2}\)\cosh\(\frac{\sqrt 2 \pi p}{n}\) \) \chi(\tau, z)_{Q=0}^p \nn\\
&~~~~~ +\(1+e^{\frac{in \pi}2} - e^{\frac{i\pi}{2n}}\(-1+e^{\frac{in\pi}2}\)\cosh\(\frac{\sqrt 2 \pi p}{n}\) \)\chi(\tau,z)_{Q=1}^p\Bigg]
\end{align}

We can do the same with the remaining two massless characters. We get
\begin{align}
\chi_{\pm 1}(ST^nS\tau,~&ST^nSz) =e^{\frac{2\pi i n z^2}{n\tau-1}} \Bigg[\frac12\(\chi_{\pm 1}(\tau,z) - \chi_{\mp 1}(\tau,z)\) \nn\\
& - \frac{1}{2\sqrt{-i n}}\int_0^\infty e^{-\frac{i\pi p^2}n}\((e^{\frac{in\pi}2}-1)\chi(\tau,z)_{Q=0}^p + (e^{\frac{in\pi}2}+1)\chi(\tau,z)_{Q=1}^p\)\Bigg]
\label{eq:masslessModular}
\end{align}
None of the terms in (\ref{eq:masslessModular}) grow exponentially. Therefore they will not be able to cancel the 
\be
(-1)^j e^{\pi \sqrt j}
\ee
growth in the vacuum term.

\section{Poincar\'e sum of minimal model characters}
\label{app:minmod}
\emph{Note added: While in the process of completing this paper, we became aware of the recent paper \cite{Meruliya:2021utr}, in which the authors computed the Poincar\'e sum of many RCFT characters and attempted to interpret the answers as averages of CFTs. In this appendix we discuss a very similar computation for the case of the minimal model characters.}

In this section, following \cite{Castro:2011zq}, we will consider the Poincar\'e sum of unitary minimal model vacuum characters. We will show that such a Poincar\'e sum \emph{cannot} be interpreted as an ensemble average of unitary minimal model CFTs. 

In \cite{Castro:2011zq}, the authors asked the question if 
\be
\sum_{\gamma \in \Gamma_{c} \backslash SL(2,\mathbb Z)} |\chi^c_{\text{vac}}(\gamma\tau)|^2  \stackrel{?}{=} Z_{\text{CFT}}(\tau, \bar\tau),
\label{eq:1111}
\ee
in other words if the Poincar\'e sum of a vacuum character at $c<1$ can be interpreted as a unitary CFT partition function\footnote{See \cite{Jian:2019ubz, Karch:2020flx} for generalizations to higher genus and boundary CFT.}. In (\ref{eq:1111}), $\Gamma_{c}$ refers to the subgroup of $SL(2,\mathbb Z)$ that leaves the vacuum character at central charge $c$ invariant. Unlike for $c\geq1$, due to the null state structure of the Virasoro algebra at $c<1$, this group will in general be a finite index subgroup of $SL(2,\mathbb Z)$ rendering the sum in (\ref{eq:1111}) finite. The authors of \cite{Castro:2011zq} showed that, up to a proportionality constant, the sum (\ref{eq:1111}) only matches the CFT partition function at $c=\frac12$ and $c=\frac7{10}$; at higher values of $c$, the Poincar\'e sum is no longer proportional to any physical CFT partition function.
For example, \cite{Castro:2011zq} showed that at $c=\frac45$, the Poincar\'e sum gives
\be
\sum_{\gamma \in \Gamma_{c} \backslash SL(2,\mathbb Z)} |\chi^{(5,6)}_{\text{vac}}(\gamma\tau)|^2  \propto \frac45 Z^{A}(\tau, \bar\tau) + \frac15 Z^{D}(\tau, \bar\tau),
\label{eq:1112}
\ee
where $Z^{A}$ and $Z^{D}$ are the two physical CFT partition functions at $c=\frac45$, coming from the $A$-series and $D$-series modular-invariant combination of characters (the tetracritical Ising model and the critical three-state Potts model, respectively). Given (\ref{eq:1112}), it is natural to ask if this sum can be interpreted as an average of CFTs at $c=\frac45$. We will show in this appendix that this cannot be the case in general, because for a generic minimal model, the modular sum of the vacuum character cannot be written as a linear combination of physical CFT partition functions. To do this, we first review the salient facts about the ADE classification of minimal models \cite{Cappelli:1986hf, Cappelli:1987xt}.

The unitary minimal models are labeled by a pair of consecutive integers $(p, p+1)$ with $p=3,4,5,\cdots$. The central charge is given by $c=1-\frac{6}{p(p+1)}$. There will be $\frac{p(p-1)}2$ different characters in the CFT, transforming as a finite-dimensional representation of $SL(2,\mathbb Z)$. There are only a finite number of modular invariant combinations of these characters with non-negative integer coefficients. For all $p$, there is an $A$-series (the diagonal invariant); for all $p \geq 5$ there is an additional $D$-series invariant; finally for $p=11, 12, 17, 18, 29, 30$ there is an exceptional $E$-series invariant. The explicit modular-invariant combination of characters for each of these partition functions can be found in \cite{Cappelli:1986hf, Cappelli:1987xt, Kato:1987td}.

However, if we relax the condition that their coefficients are non-negative integers, we will find many more modular invariants in general \cite{Cappelli:1986hf, Cappelli:1987xt, Kato:1987td}\footnote{See also the appendices of \cite{Castro:2011zq}.}. These additional modular invariants do not correspond to unitary CFTs, but nonetheless are mathematical functions that are sesquilinear combinations of the characters invariant under modular transformation. The first $p$ in which ``unphysical" modular invariants show up is at $(p,p+1)=(14,15)$ (which corresponds to $c=\frac{34}{35}$). The Poincar\'e sum of the Virasoro vacuum character $(14,15)$ \emph{cannot} be written as a linear combination of the two physical CFT partition functions (the $A$- and $D$-series), but rather has support on these unphysical modular invariants:
\be
\sum_{\gamma \in \Gamma_{c} \backslash SL(2,\mathbb Z)} |\chi^{(14,15)}_{\text{vac}}(\gamma\tau)|^2  \propto \frac8{13}Z^{A}(\tau, \bar\tau) + \frac5{13} Z^{D}(\tau, \bar\tau) - \frac1{7} X^{(1,3)}(\tau,\bar\tau) + \frac5{91}X^{(2,3)}(\tau,\bar\tau)
\label{eq:1415}
\ee
where $X^{(1,3)}(\tau,\bar\tau)$ and $X^{(2,3)}(\tau,\bar\tau)$ are the two unphysical modular invariants at $c=\frac{34}{35}$. We follow the conventions of Appendix B.3 of \cite{Castro:2011zq} in defining $X^{(1,3)}, X^{(2,3)}$. (In fact the sum in (\ref{eq:1415}) does not even have a positive expansion in the characters.)

Since the sum (\ref{eq:1415}) cannot be written as a linear combination of the only two unitary CFTs at $c=\frac{34}{35}$, we conclude that it cannot be interpreted as an averaged CFT partition function. Note that for the values of $p$ where there is no unphysical partition function, the Poincar\'e sum can be written as a linear combination of the physical CFT partition functions. For completeness, we record the answers in eq. (\ref{eq:1113}) for $p=3, 4, \cdots 13$ below:
\begin{align}
\sum_{\gamma \in \Gamma_{c} \backslash SL(2,\mathbb Z)} |\chi^{(3,4)}_{\text{vac}}(\gamma\tau)|^2  &\propto Z^{A}(\tau, \bar\tau) \nn\\
\sum_{\gamma \in \Gamma_{c} \backslash SL(2,\mathbb Z)} |\chi^{(4,5)}_{\text{vac}}(\gamma\tau)|^2  &\propto Z^{A}(\tau, \bar\tau) \nn\\
\sum_{\gamma \in \Gamma_{c} \backslash SL(2,\mathbb Z)} |\chi^{(5,6)}_{\text{vac}}(\gamma\tau)|^2  &\propto \frac45 Z^{A}(\tau, \bar\tau) + \frac15 Z^{D}(\tau, \bar\tau) \nn\\
\sum_{\gamma \in \Gamma_{c} \backslash SL(2,\mathbb Z)} |\chi^{(6,7)}_{\text{vac}}(\gamma\tau)|^2  &\propto \frac45 Z^{A}(\tau, \bar\tau) + \frac15 Z^{D}(\tau, \bar\tau) \nn\\
\sum_{\gamma \in \Gamma_{c} \backslash SL(2,\mathbb Z)} |\chi^{(7,8)}_{\text{vac}}(\gamma\tau)|^2  &\propto \frac12 Z^{A}(\tau, \bar\tau) + \frac12 Z^{D}(\tau, \bar\tau) \nn\\
\sum_{\gamma \in \Gamma_{c} \backslash SL(2,\mathbb Z)} |\chi^{(8,9)}_{\text{vac}}(\gamma\tau)|^2  &\propto \frac12 Z^{A}(\tau, \bar\tau) + \frac12 Z^{D}(\tau, \bar\tau) \nn\\
\sum_{\gamma \in \Gamma_{c} \backslash SL(2,\mathbb Z)} |\chi^{(9,10)}_{\text{vac}}(\gamma\tau)|^2  &\propto \frac23 Z^{A}(\tau, \bar\tau) + \frac13 Z^{D}(\tau, \bar\tau) \nn\\
\sum_{\gamma \in \Gamma_{c} \backslash SL(2,\mathbb Z)} |\chi^{(10,11)}_{\text{vac}}(\gamma\tau)|^2  &\propto \frac23 Z^{A}(\tau, \bar\tau) + \frac13 Z^{D}(\tau, \bar\tau) \nn\\
\sum_{\gamma \in \Gamma_{c} \backslash SL(2,\mathbb Z)} |\chi^{(11,12)}_{\text{vac}}(\gamma\tau)|^2  &\propto \frac13 Z^{A}(\tau, \bar\tau) + \frac13 Z^{D}(\tau, \bar\tau) + \frac13 Z^{E}(\tau, \bar\tau)  \nn\\
\sum_{\gamma \in \Gamma_{c} \backslash SL(2,\mathbb Z)} |\chi^{(12,13)}_{\text{vac}}(\gamma\tau)|^2  &\propto \frac13 Z^{A}(\tau, \bar\tau) + \frac13 Z^{D}(\tau, \bar\tau) + \frac13 Z^{E}(\tau, \bar\tau) \nn\\
\sum_{\gamma \in \Gamma_{c} \backslash SL(2,\mathbb Z)} |\chi^{(13,14)}_{\text{vac}}(\gamma\tau)|^2  &\propto \frac8{13} Z^{A}(\tau, \bar\tau) + \frac5{13} Z^{D}(\tau, \bar\tau).
\label{eq:1113}
\end{align}

\bibliographystyle{JHEP}
\bibliography{refmain}


\end{document}